\documentclass[prd,preprintnumbers,nofootinbib]{revtex4} 
\usepackage{graphicx} 
\usepackage{amsmath}
\usepackage{amsfonts,amsbsy}
\usepackage{amssymb}

\def\gsim{ \,\, \vcenter{\hbox{$\buildrel{\displaystyle >}\over\sim$}}
 \,\,}

\def\be{\begin{equation}}
\def\ee{\end{equation}}
\def\bea{\begin{eqnarray}}
\def\eea{\end{eqnarray}}

\begin{document}

\title{\bf Two-particle correlations in high energy collisions
  and the gluon four-point function}

\preprint{BCCUNY-HEP/10-01}
\preprint{RBRC-829}

\author{Adrian Dumitru$^{a,b,c}$, Jamal Jalilian-Marian$^{b,c}$}
\affiliation{
$^a$ RIKEN BNL Research Center, Brookhaven National
  Laboratory, Upton, NY 11973, USA\\
$^b$ Department of Natural Sciences, Baruch College, CUNY,
17 Lexington Avenue, New York, NY 10010, USA\\
$^c$ The Graduate School and University Center, City
  University of New York, 365 Fifth Avenue, New York, NY 10016, USA}

\begin{abstract}
  We derive the rapidity evolution equation for the gluon four-point
  function in the dilute regime and at small $x$ from the JIMWLK
  functional equation. We show that beyond leading order in $N_c$ the
  mean field (Gaussian) approximation where the four point function is
  factorized into a product of two point functions is violated. We
  calculate these factorization breaking terms and show that they
  contribute at leading order in $N_c$ to correlations of two produced
  gluons as a function of their relative rapidity and azimuthal angle,
  for generic (rather than back-to-back) angles. Such two-particle
  correlations have been studied experimentally at the BNL-RHIC
  collider and could be scrutinized also for $pp$ (and, in the future,
  also $AA$) collisions at the CERN-LHC accelerator.
\end{abstract}

\maketitle

\section{Production of two correlated particles}

The evolution of QCD amplitudes with energy is described by the
Balitsky hierarchy~\cite{Balitsky:1995ub} or, equivalently, by the
JIMWLK~\cite{jimwlk} functional renormalization group equations. They
essentially represent generalizations of the well-known BFKL
equation~\cite{bfkl} for the evolution of the two-point function to
evolution equations for arbitrary $n$-point functions including the non-linear
effects due to high gluon density. In the
unitarity limit of high parton density the Balitsky hierarchy is not
closed: the derivative of any $n$-point function with respect to
energy (or rapidity $Y\sim\log E$) involves all $m$-point functions
($m\ge n$). In the dilute regime, however, the hierarchy can be
truncated to obtain closed evolution equations for each $n$-point
function.

Prior work in this field has mostly focused on the evolution of the
two-point function and its perturbative unitarization at high
energies. The purpose of this paper is to point out that information
on the four-point function could be obtained from two-particle
correlations in inelastic high-energy collisions in a certain
kinematic regime (see below). Moreover, we argue that the B-JIMWLK
equation for the four point function can not be factorized as a 
product of two BFKL two-point functions. We show that the terms that
violate this factorization actually contribute to the correlation function at
leading order in $N_c$.

\begin{figure}[htb]
\begin{center}
\includegraphics[width=7cm]{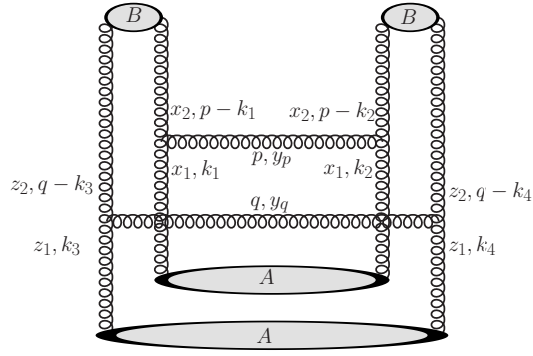}
\end{center}
\caption[a]{Correlated production of two particles with generic
  relative azimuthal angle at leading order. The blobs denote the
  unintegrated gluon distribution of the projectile $A$ or target $B$,
 respectively, and the light-cone momenta are
 $x_{1,2}=(p/\sqrt{s})\exp(\pm y_p)$,  $z_{1,2}=(q/\sqrt{s})\exp(\pm y_q)$.}
\label{fig:TwoPointC}
\end{figure}

We consider the correlation of two particles with transverse momenta
$p_\perp$, $q_\perp$ (we shall drop the subscript $\perp$ from now on
to avoid cluttering of notation) and rapidities $y_p$, $y_q$,
respectively:
\be
C({\bf p},{\bf q}) = 
 \left\langle \frac{dN_2}{d^2pdy_p\,d^2qdy_q} \right\rangle
 -
 \left\langle \frac{dN}{d^2pdy_p} \right\rangle
 \left\langle \frac{dN}{d^2qdy_q} \right\rangle ~.
\ee
The brackets denote an average over events and the momentum
distributions shall be normalized according to
\bea
\int d^2pdy_p \, 
  \left\langle\frac{dN}{d^2pdy_p}\right\rangle &=& \langle N\rangle~, \\
\int d^2pdy_p d^2qdy_q \, 
  \left\langle\frac{dN_2}{d^2pdy_p\,d^2qdy_q}\right\rangle
&=& \langle N^2\rangle~,
\eea
where $\langle N\rangle$ is the total average multiplicity per event.
It has been argued in ref.~\cite{Dumitru:2008wn} that in the
high-energy limit (but fixed $p$, $q$, $y_p$, $y_q$) the leading
contribution to $C({\bf p},{\bf q})$ is due to diagrams such as the
one depicted in fig.~\ref{fig:TwoPointC}. For these diagrams the hard
amplitudes are disconnected but the correlations arise because for
either one (or both) of the colliding hadrons the ladders in the amplitude
and/or the conjugate amplitude connect to the same color source. These
two-point functions are essentially the unintegrated gluon
distributions of the hadrons; they are of order $1/g^2$ when the
transverse momentum in the ladder is below the saturation momentum
$Q_s$ of the corresponding hadron.

Diagrams such as fig.~\ref{fig:TwoPointC} should dominate $C(p,q)$
even at high (but not asymptotically high) transverse momentum,
$p,q\gsim Q_s$, provided one considers {\em generic} relative angles
$\cos\phi\equiv p\cdot q/(|p|\, |q|)$ (in particular, away from the
region of ``back-to-back'' jets, $\phi\simeq \pi$). On the other hand,
at leading order in $\alpha_s$, when $p,q\gg Q_s$ the gluon pair
should originate from the same ladder; when the rapidity difference between
the two produced gluons and the two beams are smaller than
$\sim1/\alpha_s$ the ladder is DGLAP-ordered and $C(p,q)$ should
approach $\delta(p+q)$ (back-to-back dijet). When $|y_p-y_q|\gsim
1/\alpha_s$ the delta-function is smeared out by a BFKL-ordered ladder
inbetween the produced gluons (Mueller-Navelet
jets~\cite{Mueller:1986ey}). Instead, here we consider the situation
where $p,q$ are somewhat larger than but on the order of $Q_s$; also,
$|y_p-y_q|$ should be significantly smaller than the total rapidity
window between the two beams, and the relative azimuthal angle
$\phi\ll\pi$. When $p$ and $q$ are on the order of a few GeV it is
necessary to subtract the background of uncorrelated particle pairs to
reveal the structure of the correlation function.

We note that two-particle correlations away from the back-to-back
regime have recently been measured at the BNL-RHIC accelerator at
$\sqrt{s}=200$~GeV (per colliding nucleon pair) for proton-proton,
deuteron-gold, and gold-gold
collisions~\cite{ridge_star,ridge_phobos,ph:2008cqb}. For the former
systems only a narrow peak due to fragmentation of the triggered
parton have been observed. For collisions of heavy ions, on the other
hand, $C(p,q)$ exhibits a ``ridge''-like structure: it is narrow in
$\phi$ but extends over several units in $\Delta y = |y_p-y_q|$. The
absence of measurable correlations in $pp$ and $d+Au$ collisions may
be due to the smallness of the saturation momentum $Q_s$ for a proton
or deuteron at RHIC energy. Also, the measurements from RHIC might be
expected to be rather sensitive to the initial conditions for the
evolution equation at moderately small $x_0$. At the higher energies
of CERN's LHC collider, the saturation momentum of a proton measured
from the central rapidity region is expected to be on the order of
1~GeV and such correlations could be sufficiently strong to provide
information about the QCD four-point function at small $x$.

The diagrams like the one from fig.~\ref{fig:TwoPointC} arise from
factorization of the four-point functions in the field of the
projectile/target into products of two-point
functions~\cite{Dumitru:2008wn} (unintegrated gluon
distributions). Doing so, however, picks up only the leading-$N_c$
contribution to the four-point function. More generally, $C({\bf
  p},{\bf q})$ is given by
\bea
\left\langle \frac{dN_2}{d^2pdy_p\,d^2qdy_q} \right\rangle
 &=&  \frac{g^{12}}{64 (2\pi)^6}\, 
\left(
f_{g aa^\prime} f_{g^\prime b b^\prime}
f_{gcc^\prime} f_{g^\prime dd^\prime} 
\right)
\int \prod_{i=1}^4 \frac{d^2 k_{i}}{(2\pi)^2 k_{i}^2}
\frac{L_\mu(p,k_{1})L^\mu(p,k_{2})}{(p-k_{1})^2(p-k_{2})^2}
\frac{L_\nu(q,k_{3})L^\nu(q,k_{4})}{(q-k_{3})^2 (q-k_{4})^2} \nonumber\\
& & \quad \times 
\left< {\rho^*}_A^a (k_{2}) {\rho^*}_A^{b} (k_{4}) 
{{\rho}_A}^c (k_{1}) {{\rho}_A}^d (k_{3})\,
{\rho^*}_B^{a^\prime} (p-k_{2})
{\rho^*}_B^{b^\prime} (q-k_{4})
{{\rho}_B}^{c^\prime}(p-k_{1}) 
{{\rho}_B}^{d^\prime} (q-k_{3})\right> \\
& =&
\frac{g^{12}}{64 (2\pi)^6}\, 
\left(
f_{g aa^\prime} f_{g^\prime b b^\prime}
f_{gcc^\prime} f_{g^\prime dd^\prime} 
\right)
\int \prod_{i=1}^4 \frac{d^2 k_{i}}{(2\pi)^2 k_{i}^2}
\frac{L_\mu(p,k_{1})L^\mu(p,k_{2})}{(p-k_{1})^2(p-k_{2})^2}
\frac{L_\nu(q,k_{3})L^\nu(q,k_{4})}{(q-k_{3})^2 (q-k_{4})^2} \nonumber\\
& & \quad \times 
\left< {\rho^*}_A^a (k_{2}) {\rho^*}_A^{b} (k_{4}) 
{{\rho}_A}^c (k_{1}) {{\rho}_A}^d (k_{3})\right> \left<
{\rho^*}_B^{a^\prime} (p-k_{2})
{\rho^*}_B^{b^\prime} (q-k_{4})
{{\rho}_B}^{c^\prime}(p-k_{1}) 
{{\rho}_B}^{d^\prime} (q-k_{3})\right> \label{eq:Cpq_FourPoint}
\eea
In the second step we have assumed factorization of the wave functions
of projectile and target. $L^\mu$ denotes the Lipatov vertex which
satisfies
\bea
L_\mu(p,k_1)L^\mu(p,k_2) &=& -\frac{4}{p^2} 
\left[\delta^{ij}\delta^{nm} + \epsilon^{ij}\epsilon^{nm}\right] 
k_1^i (p-k_1)^j \, k_2^n (p-k_2)^m  \\
L_\mu(p,k)L^\mu(p,k) &=& -\frac{4k^2}{p^2} \left(p-k\right)^2~.
\eea

\begin{figure}[htb]
\begin{center}
\includegraphics[width=7cm]{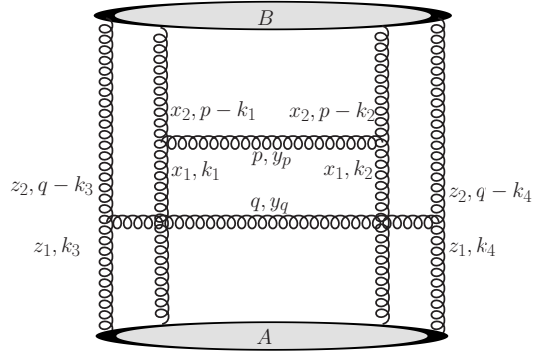}
\end{center}
\caption[a]{Correlated production of two particles with generic
  relative azimuthal angle at leading order. The blobs denote the
  four-point functions for the projectile $A$ or target $B$,
 respectively.}
\label{fig:FourPointC}
\end{figure}

\begin{figure}[htb]
\begin{center}
\includegraphics[width=7cm]{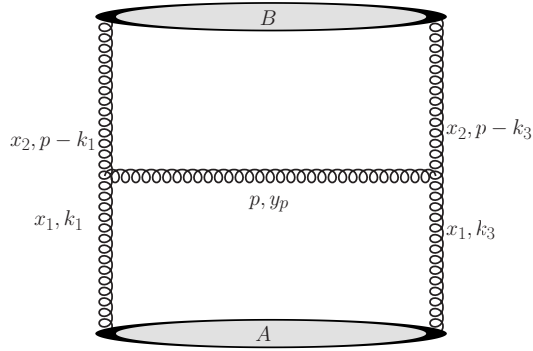}
\end{center}
\caption[a]{Single-particle production from
  $k_\perp$-factorization at leading order. The blobs denote the
  unintegrated gluon distribution of the projectile $A$ or target $B$,
 respectively.}
\label{fig:Single}
\end{figure}

The expression~(\ref{eq:Cpq_FourPoint}) is depicted in
fig.~\ref{fig:FourPointC}.
Here, $\rho(r)$ denotes the color charge density per unit transverse
area at a transverse coordinate $r$ and $\rho(k)$ is its Fourier
transform. Its two-point function is related to the unintegrated
gluon distribution $\Phi(x,k^2)$ via
\be \label{eq:Phi_TwoPoint}
\left< {\rho^*}^a(k) \rho^b(k')\right>(x) = \frac{1}{\alpha_s} 
   \frac{\delta^{ab}}{N_c^2-1} (2\pi)^3 \delta(k-k') \, \Phi(x,k^2)~.
\ee
With this normalization one recovers the LO $k_\perp$-factorization
formula for the single-inclusive distribution from the
diagram~\ref{fig:Single} with the standard prefactor~\cite{Gribov:1984tu}:
\be
\frac{dN}{d^2p dy} = 4\alpha_s \frac{N_c}{N_c^2-1}
\frac{\sigma_0}{p^2} \int d^2k \frac{\Phi_A(x_1,k^2)}{k^2} 
\frac{\Phi_B(x_2,(p-k)^2)}{(p-k)^2}~,
\ee
where $\sigma_0$ is the transverse area of the collision (note that in
our convention $\Phi(x,k^2)$ is the density of gluons per unit
transverse area and it therefore contains a factor of $1/\sigma_0$).

In a mean field (and large $N_c$) approximation one may factorize the
four-point functions from eq.~(\ref{eq:Cpq_FourPoint}) into products
of two-point functions,
\be \label{eq:4point_fact_2point}
\left< \rho^a \rho^b \rho^c \rho^d \right> =
 \delta^{ab} \delta^{cd} (\rho^2)^2 + 
 \delta^{ac} \delta^{bd} (\rho^2)^2 + 
 \delta^{ad} \delta^{bc} (\rho^2)^2 + \cdots~,
\ee
where $\rho^2\equiv \langle\rho\rho\rangle$, and the momentum
dependence of the two-point function has been suppressed. Then, one of
the nine contractions corresponds to the square of the
single-inclusive distribution: contract the first $\rho$ with the
third and the second with the fourth, for both projectile and
target. The color factor for this diagram is\footnote{Not including
  factors of $N_c$ which will enter once $\langle\rho\rho\rangle$ is
  expressed through $\Phi$ via eq.~(\ref{eq:Phi_TwoPoint}).}
\bea
& &
f_{g aa^\prime} f_{g^\prime b b^\prime}
f_{gcc^\prime} f_{g^\prime dd^\prime} 
\left< {\rho^*}_A^a
{{\rho}_A}^c
\right>\left<
{\rho^*}_A^{b}
{{\rho}_A}^d
\right> \left<
{\rho^*}_B^{a^\prime}
{{\rho}_B}^{c^\prime}
\right>\left<{\rho^*}_B^{b^\prime}
{{\rho}_B}^{d^\prime}
\right> \\
& \sim & f_{g aa^\prime} f_{g^\prime b b^\prime}
f_{gcc^\prime} f_{g^\prime dd^\prime} 
\delta^{ac}
\delta^{bd}
\delta^{a^\prime c^\prime}
\delta^{b^\prime d^\prime}
= N_c^2 (N_c^2-1)^2~.   \label{eq:SingleSquareColor}
\eea
The remaining eight diagrams correspond to a color factor of (we take
fig.~\ref{fig:TwoPointC} as an example)
\bea
& & f_{g aa^\prime} f_{g^\prime b b^\prime}
f_{gcc^\prime} f_{g^\prime dd^\prime} 
\left< {\rho^*}_A^a
{{\rho}_A}^c
\right>\left<
{\rho^*}_A^{b}
{{\rho}_A}^d
\right> \left<
{\rho^*}_B^{a^\prime}
{\rho^*}_B^{b^\prime}
\right>\left<
{{\rho}_B}^{c^\prime}
{{\rho}_B}^{d^\prime}
\right> \\
& \sim & f_{g aa^\prime} f_{g^\prime b b^\prime}
f_{gcc^\prime} f_{g^\prime dd^\prime} 
\delta^{ac}
\delta^{bd}
\delta^{a^\prime b^\prime}
\delta^{c^\prime d^\prime}
= N_c^2 (N_c^2-1)~.   \label{eq:Fig1_Color}
\eea
Thus, two-particle correlations are suppressed by a factor of
$N_c^2-1$ as compared to uncorrelated production. For this reason, the
leading-$N_c$ {\em ansatz}~(\ref{eq:4point_fact_2point}) may not
capture the complete result for $C(p,q)$. Below, we derive the
evolution equation for the four-point function from JIMWLK.  We
determine the corrections beyond the mean-field and large-$N_c$
approximations to the rhs of~(\ref{eq:4point_fact_2point})
and show that these corrections contribute at the same order in $N_c$
to the correlation function.

In this regard, we should point out that $N_c$ corrections to the
two-point function in the dense regime were found to be exceptionally
small~\cite{Kovchegov:2008mk}. However, this needs not be true for the
four-point function. In fact, we shall argue below that we do not
expect $N_c$ corrections to the four-point function to be anomalously
small, even in the dilute regime. A verification or falsification of
this expectation via exact numerical solutions would be very valuable.

\section{Evolution equation for the four-point function}

In this section we present the equation describing the rapidity
evolution of the four-point function $\left< \alpha^a_r \alpha^b_{\bar
    r} \alpha^c_s \alpha^d_{\bar s} \right>$ obtained from the JIMWLK
equations, which include terms of subleading order in $N_c$. In this
context it is more natural to work in coordinate space, so $r$, $s$,
$\bar r$, $\bar s$ denote transverse coordinates; the four-point
function in momentum space can be obtained by Fourier transform. We
also find it preferable to work with the fields $\alpha$ rather than
the color charge densities $\rho$; at leading order and in covariant
gauge, they are related in coordinate space by
\be
A^\mu(x^+,r) \equiv \delta^{\mu -} \alpha(x^+,r) = 
     - g \, \delta^{\mu -} \delta(x^+) \frac{1}{\nabla^2_\perp} \rho(x^+,r)~,
\ee
for a hadron moving at the speed of light in the negative
$z$-direction. Since this field also satisfies $A^+=0$, the only
non-vanishing field-strength is $F^{-i} = - \partial^i \alpha$.
In momentum space we have the relation $k^2\alpha(k)=g\rho(k)$.

The JIMWLK evolution equation for the four-point function to lowest order
in the fields can be shown to be (see appendix)
\bea
\frac{d}{dY} 
\langle \alpha_r^a \alpha_{\bar r}^b \alpha_s^c \alpha_{\bar s}^d \rangle & =&
\frac{g^2N_c}{(2\pi)^3} \int d^2z
\left<
  \frac{\alpha_z^a \alpha_{\bar r}^b \alpha_s^c \alpha_{\bar s}^d}{(r-z)^2} +
  \frac{\alpha_r^a \alpha_z^b \alpha_s^c \alpha_{\bar s}^d}{(\bar{r}-z)^2} +
  \frac{\alpha_r^a \alpha_{\bar r}^b \alpha_z^c \alpha_{\bar s}^d}{(s-z)^2} +
  \frac{\alpha_r^a \alpha_{\bar r}^b \alpha_s^c \alpha_z^d}{(\bar{s}-z)^2}
  -4 \frac{\alpha_r^a \alpha_{\bar r}^b \alpha_s^c \alpha_{\bar s}^d}{z^2}
\right> \nonumber\\
&& \hspace{-2.2cm}
+ \frac{g^2}{\pi} \int \frac{d^2z}{(2\pi)^2} \left<
 f^{e\kappa a}f^{f\kappa b} \frac{(r-z)\cdot(\bar r-z)}{(r-z)^2 (\bar r-z)^2} 
 \left[\alpha_r^e\alpha_{\bar r}^f - \alpha_r^e\alpha_z^f - 
       \alpha_z^e\alpha_{\bar r}^f + \alpha_z^e\alpha_z^f \right]
       \alpha_s^c \alpha_{\bar s}^d \right. \nonumber\\
& & +
 f^{e\kappa a}f^{f\kappa c} \frac{(r-z)\cdot(s-z)}{(r-z)^2 (s-z)^2} 
 \left[\alpha_r^e\alpha_s^f - \alpha_r^e\alpha_z^f - 
       \alpha_z^e\alpha_s^f + \alpha_z^e\alpha_z^f \right]
       \alpha_{\bar r}^b \alpha_{\bar s}^d \nonumber\\
& & +
 f^{e\kappa a}f^{f\kappa d} \frac{(r-z)\cdot(\bar s-z)}{(r-z)^2 (\bar s-z)^2} 
 \left[\alpha_r^e\alpha_{\bar s}^f - \alpha_r^e\alpha_z^f - 
       \alpha_z^e\alpha_{\bar s}^f + \alpha_z^e\alpha_z^f \right]
       \alpha_{\bar r}^b \alpha_s^c \nonumber\\
& & +
 f^{e\kappa b}f^{f\kappa c} \frac{(\bar r-z)\cdot(s-z)}{(\bar r-z)^2 (s-z)^2} 
 \left[\alpha_{\bar r}^e\alpha_s^f - \alpha_{\bar r}^e\alpha_z^f - 
       \alpha_z^e\alpha_s^f + \alpha_z^e\alpha_z^f \right]
       \alpha_r^a \alpha_{\bar s}^d \nonumber\\
& & +
 f^{e\kappa b}f^{f\kappa d} \frac{(\bar r-z)\cdot(\bar s-z)}{(\bar r-z)^2 (\bar s-z)^2} 
 \left[\alpha_{\bar r}^e\alpha_{\bar s}^f - \alpha_{\bar r}^e\alpha_z^f - 
       \alpha_z^e\alpha_{\bar s}^f + \alpha_z^e\alpha_z^f \right]
       \alpha_r^a \alpha_s^c \nonumber\\
& & + \left.
 f^{e\kappa c}f^{f\kappa d} \frac{(s-z)\cdot(\bar s-z)}{(s-z)^2 (\bar s-z)^2} 
 \left[\alpha_s^e\alpha_{\bar s}^f - \alpha_s^e\alpha_z^f - 
       \alpha_z^e\alpha_{\bar s}^f + \alpha_z^e\alpha_z^f \right]
       \alpha_r^a \alpha_{\bar r}^b \right>~.  \label{eq:4point_Evol}
\eea
This expression neglects contributions from higher $n$-point functions
on the rhs; in the dilute regime, i.e.\ when the transverse momenta of
the produced particles are higher than the saturation momenta of the
colliding hadrons, this approximation should be justified.

In order to derive the color structure of corrections beyond the
large-$N_c$ approximation, we factorize the product of four point
functions on the rhs of eq.~(\ref{eq:4point_Evol}) into products of
two point functions. This Gaussian approximation reduces the evolution
equation for the four point function to a product of two BFKL
equations (for the two point function) plus extra terms which provide
corrections to the factorization~(\ref{eq:4point_fact_2point}). The
result is

\be
\frac{d}{dY} 
\langle \alpha_r^a \alpha_{\bar r}^b \alpha_s^c \alpha_{\bar s}^d \rangle =
\frac{d}{dY} \left[
    \delta^{ac} \delta^{bd} \alpha^2_{\bar r-\bar s} \alpha^2_{r-s} +
    \delta^{ab} \delta^{cd} \alpha^2_{s-\bar s} \alpha^2_{r-\bar r} + 
    \delta^{ad} \delta^{bc} \alpha^2_{r-\bar s} \alpha^2_{\bar r-s}
   \right]
- \frac{\alpha_s}{2\pi^2} \int d^2z \left[F_0^{abcd} + F_1^{abcd} + F_2^{abcd}\right]
\label{eq:fact_correction}
\ee
where 
\bea
F_0^{abcd} &\equiv& f^{a\kappa b}f^{c\kappa d}
    \frac{(r-s)^2}{(r-z)^2 (s-z)^2} \alpha^2_{r-\bar r} \alpha^2_{s-\bar s}
+ f^{a\kappa d}f^{b\kappa c}
 \left[
   \frac{(r-\bar r)^2}{2(r-z)^2 (\bar r-z)^2} - 
   \frac{(r-s)^2}{2(r-z)^2 (s-z)^2} +
   (r\leftrightarrow s, \bar s\leftrightarrow \bar r)
 \right] \alpha^2_{r-\bar s} \alpha^2_{\bar r - s} \nonumber \\
F_1^{abcd} &\equiv&  f^{a\kappa b}f^{c\kappa d} \left[~
      \left( 
        \frac{1}{(r-z)^2} - \frac{(s-r)^2}{(r-z)^2(s-z)^2} 
      \right)
       \alpha^2_{r-\bar r} \alpha^2_{z-\bar s} + 
       (r\leftrightarrow \bar s, s\leftrightarrow \bar r) \right] \nonumber\\
&+ &   f^{a\kappa d}f^{b\kappa c} \left[~
      \left( 
        \frac{1}{(r-z)^2} - \frac{(\bar r-r)^2}{(r-z)^2(\bar r-z)^2} 
        - \frac{1}{(\bar s-z)^2} + \frac{(\bar s-\bar r)^2}{(\bar r-z)^2(\bar s-z)^2} 
      \right)
        \alpha^2_{r-\bar s} \alpha^2_{z-s}  +
      (r\leftrightarrow s, \bar s\leftrightarrow \bar r) \right] \nonumber \\
F_2^{abcd} &\equiv& f^{a\kappa b}f^{c\kappa d} \left[\left[\left( \frac{(r-s)^2}{(r-z)^2 (s-z)^2} - \frac{1}{(r-z)^2} 
 - \frac{1}{(s-z)^2} \right) \alpha^2_{z-\bar r} \alpha^2_{z-\bar s} -   (s\leftrightarrow \bar s) \right] - 
 (r\leftrightarrow \bar r) \right].
\label{eq:Fs}
\eea
In (\ref{eq:Fs}) all terms in $F_0$, $F_1$ and $F_2$ are to be
duplicated with the substitutions indicated explicitly in the
brackets. Second, all terms in $F_0$ and $F_1$ are to be duplicated
again substituting $a\leftrightarrow b$ and $r\leftrightarrow \bar
r$. Then, all terms in $F_0$ and $F_1$ should be duplicated a third
time exchanging $c\leftrightarrow d$ and $s\leftrightarrow \bar
s$. Furthermore, all terms in $F_2$ are to be duplicated while letting
$b\leftrightarrow c$ and $r\leftrightarrow \bar s$. Finally, the terms
obtained in the last substitution (only) in $F_2$ should be duplicated
exchanging $c\leftrightarrow d$ and $r\leftrightarrow \bar r$.

The first term in (\ref{eq:fact_correction}) provides the
leading-$N_c$ contribution to the four-point function. The second term
gives corrections beyond the large-$N_c$
factorization~(\ref{eq:4point_fact_2point}). Since an analytic
solution to the evolution equation for the four point function is not
within our reach, a numerical investigation of these terms and their
magnitude would be extremely useful. Nevertheless, from
\be
\partial_Y \left< \rho^a \rho^b \rho^c \rho^d \right> \sim
\alpha_s N_c \delta^{ab}\delta^{cd} (\rho^2)^2 +
\alpha_s f^{ac\kappa} f^{bd\kappa} (\rho^2)^2~~~,~~~
\mathrm{with}~\rho^2(Y) \sim e^{\alpha_sN_cY}
\ee
one might expect
that, generically, the solution to this equation has the following
color structure:
\bea
\left< \rho^a \rho^b \rho^c \rho^d \right> &\sim&
 \delta^{ab} \delta^{cd} (\rho^2)^2 + 
 \delta^{ac} \delta^{bd} (\rho^2)^2 + 
 \delta^{ad} \delta^{bc} (\rho^2)^2 + \label{eq:4point_leading}\\
& &
 \frac{1}{N_c}f^{ac\kappa}f^{bd\kappa} (\rho^2)^2 +
 \frac{1}{N_c}f^{ab\kappa}f^{cd\kappa} (\rho^2)^2 +
 \frac{1}{N_c}f^{ad\kappa}f^{bc\kappa} (\rho^2)^2~. \label{eq:4point_subleading}
\eea
(Note that the various two-point functions depend on different
coordinates/momenta and so each of the above terms is distinct.) The
color factors emerging from the products of the Kronecker tensors have
already been discussed above, eqs.~(\ref{eq:SingleSquareColor})
and~(\ref{eq:Fig1_Color}). However, some of the products of a
leading-$N_c$ term from the first line~(\ref{eq:4point_leading}) with
a subleading-$N_c$ term from the second line~(\ref{eq:4point_subleading})
also contribute at the same order $N_c^2(N_c^2-1)$. For example,
\be
\frac{1}{N_c} \delta^{a'c'} \delta^{b'd'} 
  f_{g aa^\prime} f_{g^\prime b b^\prime} f_{gcc^\prime} f_{g^\prime dd^\prime}\,
  f^{ab\kappa}f^{cd\kappa} =
  N_c \delta^{ac} \delta^{bd}\, f^{ab\kappa}f^{cd\kappa} = N_c^2(N_c^2-1) ~.
\ee
The other products are worked out in appendix~\ref{sec:ColorFactorsApp}.

This shows that some of the subleading-$N_c$ contributions from the
four-point function actually enter $C(p,q)$ at leading order, compare
to eq.~(\ref{eq:Fig1_Color}). Previous results from the
literature~\cite{Dumitru:2008wn} (also see~\cite{Dusling:2009ni}) are
therefore not complete. Nevertheless, the correlations described
here should still extend over several units in
$|y_p-y_q|$~\cite{Dusling:2009ni}. Quantitative results for the JIMWLK
four-point function and for the corresponding two-particle correlations
$C({\bf p},{\bf q})$ as functions of the transverse momenta $p$, $q$,
relative azimuth $\phi$ and relative rapidity $|y_p-y_q|$ remain to be
found.

In summary, we have argued that two-particle correlations from
high-energy collisions may provide some insight into the QCD
four-point function. This should be the case, in particular, when the
transverse momenta of the produced particles are not very much higher
than the saturation momenta of the colliding hadrons and when their
relative azimuthal angle is sufficiently less than $\pi$. The narrow
(in both azimuthal and polar angle) jet-like fragmentation peak should
sit on top of a ``background'' which is broader in the relative
rapidity $|y_p-y_q|$.

If expanded in powers of $N_c$, the leading contribution to the
four-point function is given by the product of two BFKL two-point
functions. However, we find that genuine B-JIMWLK subleading-$N_c$
corrections also appear in the correlation function $C({\bf p},{\bf
  q})$, at leading non-vanishing order in $N_c$. The correlations
mentioned here represent an interesting opportunity to study the
non-trivial structure of the four-point function of the B-JIMWLK
hierarchy, both theoretically and experimentally.

\appendix

\section{Two-point function and BFKL}
The JIMWLK equation for the two-point function is
\bea
\frac{d}{dY} \langle \alpha_r^a \alpha_s^c \rangle &=&
  \frac{1}{2} \int d^2x d^2y \frac{\delta}{\delta\alpha_x^b}
     \eta^{bd}_{xy} \frac{\delta}{\delta\alpha_y^d} 
     \alpha_r^a \alpha_s^c \\
 &=& \frac{1}{2} \int d^2x d^2y \left[ \frac{\delta}{\delta\alpha_x^b}
     \eta^{bd}_{xy} \right] \frac{\delta}{\delta\alpha_y^d} 
     \alpha_r^a \alpha_s^c
  ~+~ \frac{1}{2} \int d^2x d^2y \, \eta^{bd}_{xy} 
     \frac{\delta}{\delta\alpha_x^b} \frac{\delta}{\delta\alpha_y^d} 
     \alpha_r^a \alpha_s^c ~,   \label{eq:2}
\eea
where
\be  \label{eq:9}
\eta^{bd}_{xy} = \frac{1}{\pi} \int \frac{d^2z}{(2\pi)^2}
     \frac{(x-z)\cdot(y-z)}{(x-z)^2 (y-z)^2} \left[
       1 + V^\dagger_x V_y - V^\dagger_x V_z - V^\dagger_z V_y \right]^{bd}~.
\ee
We start with the first term from eq.~(\ref{eq:2}):
\bea
\frac{\delta}{\delta\alpha_x^b} \eta^{bd}_{xy} &=& 
 \frac{1}{\pi} \int \frac{d^2z}{(2\pi)^2}
 \frac{(x-z)\cdot(y-z)}{(x-z)^2 (y-z)^2}
 \left[V^\dagger_x \frac{\delta}{\delta\alpha_x^b}V_y 
    - V^\dagger_x \frac{\delta}{\delta\alpha_x^b} V_z 
    - V^\dagger_z \frac{\delta}{\delta\alpha_x^b} V_y \right]^{bd} \label{eq:4}\\
 &=& \frac{-ig}{\pi} \int \frac{d^2z}{(2\pi)^2}
  \frac{(x-z)\cdot(y-z)}{(x-z)^2 (y-z)^2}
  \left[
    \delta(x-y) V^\dagger_x V_y t^b
   -\delta(x-z) V^\dagger_x V_z t^b
   -\delta(x-y) V^\dagger_z V_y t^b \right]^{bd}~. \label{eq:5}
\eea
Note that in (\ref{eq:4}) the derivative does not act on the
$V^\dagger$ on the left because
\be
\left(\left[\frac{\delta}{\delta\alpha_x^b} V^\dagger_u\right]X\right)_{bd} 
= ig\delta(x-u) \left(t^b V^\dagger_u X\right)_{bd} 
\sim (t^b)_{be} \left(V^\dagger_u X\right)_{ed} = 0~.
\ee

The first term in the bracket from eq.~(\ref{eq:5}) vanishes because
$\delta(x-y) V^\dagger_x V_y t^b = \delta(x-y) t^b$ and $(t^b)_{bd}=0$.
To apply the same argument to the second term from~(\ref{eq:5}) we
rewrite
\be
2 (x-z)\cdot(y-z) = (x-z)^2 + (y-z)^2 -[(x-z)-(y-z)]^2 =
(x-z)^2 + (y-z)^2 - (x-y)^2~.
\ee
Hence, the second term from~(\ref{eq:5}) becomes
\be
\frac{ig}{2\pi} \int \frac{d^2z}{(2\pi)^2} 
\frac{(x-z)^2 + (y-z)^2 - (x-y)^2}{(x-z)^2 (y-z)^2}\delta(x-z)
\left[V^\dagger_x V_z t^b\right]^{bd} = 0~.
\ee
The first term again vanishes when $x=z$ while for the other two terms
the divergences at $x=z$ cancel and so they vanish due to the color
structure.

We can therefore simplify eq.~(\ref{eq:5}) to
\bea
\frac{\delta}{\delta\alpha_x^b} \eta^{bd}_{xy} &=& 
 \frac{ig}{\pi} \delta(x-y) \int \frac{d^2z}{(2\pi)^2}
  \frac{(x-z)\cdot(y-z)}{(x-z)^2 (y-z)^2}
  \left[V^\dagger_z V_y t^b \right]^{bd}~. \label{eq:5b}
\eea

For the second term from eq.~(\ref{eq:2}) we need
\bea
\frac{\delta}{\delta\alpha_y^d} \alpha_r^a \alpha_s^c &=&
   \delta^{ad}(r-y) \alpha_s^c + \delta^{cd}(s-y) \alpha_r^a \\
\frac{\delta}{\delta\alpha_x^b} \frac{\delta}{\delta\alpha_y^d} 
     \alpha_r^a \alpha_s^c &=&
    \delta^{ad}(r-y) \delta^{bc}(x-s) + \delta^{cd}(s-y) \delta^{ab}(r-x)~.
\eea
Eq.~(\ref{eq:2}) turns into
\bea
\frac{d}{dY} \langle \alpha_r^a \alpha_s^c \rangle &=&
  \frac{ig}{(2\pi)^3} \int d^2x d^2y d^2z  
  \frac{(x-z)\cdot(y-z)}{(x-z)^2 (y-z)^2}
  \left\{ \delta(x-y) \left[V^\dagger_z V_y t^b \right]^{bd}
         \left[ \delta^{ad}(r-y) \alpha_s^c + \delta^{cd}(s-y)
           \alpha_r^a \right]
  \right\}  \\
& & \hspace{-2.5cm} +
  \frac{1}{(2\pi)^3} \int d^2x d^2y d^2z
  \frac{(x-z)\cdot(y-z)}{(x-z)^2 (y-z)^2}
  \left[1 + V^\dagger_x V_y - V^\dagger_x V_z - V^\dagger_z V_y \right]^{bd}
  \left[
    \delta^{ad}(r-y) \delta^{bc}(x-s) + \delta^{cd}(s-y) \delta^{ab}(r-x)
  \right] \\
&=&
  \frac{ig}{(2\pi)^3} \int d^2z
  \left\{ \frac{\left[V^\dagger_z V_r t^b \right]^{ba}\alpha_s^c}{(r-z)^2}
 +        \frac{\left[V^\dagger_z V_s t^b \right]^{bc}\alpha_r^a}{(s-z)^2} 
  \right\} \\
& & +
  \frac{1}{(2\pi)^3} \int d^2z
  \frac{(s-z)\cdot(r-z)}{(s-z)^2 (r-z)^2} \left\{
     \left[1 + V^\dagger_s V_r - V^\dagger_s V_z - V^\dagger_z V_r \right]^{ca}
    +\left[1 + V^\dagger_r V_s - V^\dagger_r V_z - V^\dagger_z V_s \right]^{ac}
    \right\} \\
&=&
  \frac{2ig}{(2\pi)^3} \int d^2z
  \frac{\left[V^\dagger_z V_y t^b \right]^{ba}\alpha_s^c}{(r-z)^2}
 +\frac{2}{(2\pi)^3} \int d^2z
  \frac{(s-z)\cdot(r-z)}{(s-z)^2 (r-z)^2}
  \left[1 + V^\dagger_r V_s - V^\dagger_r V_z - V^\dagger_z V_s \right]^{ac}~.
     \label{eq:16}
\eea
To expand the rhs to second order in the fields we need the following
expressions:
\bea
\left[V^\dagger_z V_y t^b \right]^{ba}\alpha_s^c &=& 
   ig\left( \alpha_z^d - \alpha_r^d \right) \left[t^dt^b\right]^{ba}\alpha_s^c
   + \cdots \\
&=& ig\left( \alpha_z^d - \alpha_r^d \right) 
      \left[t^d_{be}t^b_{ea}\right]\alpha_s^c \\
&=& ig\left( \alpha_z^d - \alpha_r^d \right) 
      \left[if^{dbe}\, if^{bea}\right]\alpha_s^c \\
&=& -ig N_c \delta^{ad} \left( \alpha_z^d - \alpha_r^d \right) \alpha_s^c \\
&=& ig N_c \left( \alpha_r^a - \alpha_z^a \right) \alpha_s^c
\eea
and
\bea
\left[1 + V^\dagger_r V_s - V^\dagger_r V_z - V^\dagger_z V_s \right]^{ac}
&=&
ig\left[ \left(\alpha_r - \alpha_s\right)
        -\left(\alpha_r - \alpha_z\right)
        -\left(\alpha_z - \alpha_s\right) \right]^{ac} \\
&& +
(ig)^2 \left[ \alpha_r^2 + \alpha_s^2 - \alpha_r\alpha_s
             -\alpha_r^2 - \alpha_z^2 + \alpha_r\alpha_z
             -\alpha_z^2 - \alpha_s^2 + \alpha_z\alpha_s
       \right]^{ac} \label{eq:23}\\
&=&
g^2 \left[\alpha_z^2 + \alpha_r\alpha_s - \alpha_r\alpha_z - \alpha_z\alpha_s
    \right]^{ac} \label{eq:24}\\
&=&
g^2 \left[\alpha_z^b\alpha_z^d + \alpha_r^b\alpha_s^d - 
          \alpha_r^b\alpha_z^d - \alpha_z^b\alpha_s^d
    \right]
    t^b_{ae} t^d_{ec} \\
&=& g^2 f^{bea} f^{dec} \left[\alpha_z^b\alpha_z^d + \alpha_r^b\alpha_s^d - 
          \alpha_r^b\alpha_z^d - \alpha_z^b\alpha_s^d \right] ~.
\eea
In the step from~(\ref{eq:23}) to~(\ref{eq:24}) we have used the
rapidity ordering of the fields; hence, only one of the $\sim
\alpha_z^2$ terms can contribute.

Using these expressions in~(\ref{eq:16}) gives
\bea
\frac{d}{dY} \langle \alpha_r^a \alpha_s^c \rangle &=& -
\frac{2g^2N_c}{(2\pi)^3} \int \frac{d^2z}{(r-z)^2}
  \left[\alpha_r^a -\alpha_z^a  \right] \alpha_s^c \nonumber \\
& + &\frac{2g^2f^{bea} f^{dec}}{(2\pi)^3} \int d^2z
  \frac{(s-z)\cdot(r-z)}{(s-z)^2 (r-z)^2}
  \left[ \alpha_z^b\alpha_z^d + \alpha_r^b\alpha_s^d - 
          \alpha_r^b\alpha_z^d - \alpha_z^b\alpha_s^d \right]~. \label{eq:27}
\eea
We now take the expectation value on the rhs using
\be \label{eq:28}
\langle \alpha_r^a \alpha_s^c \rangle = \delta^{ac} \alpha^2_{r-s}
\ee
where $\alpha^2_{r-s}$ is essentially the unintegrated gluon
distribution function. This turns eq.~(\ref{eq:27}) into (we drop an overall
$\delta^{ac}$)
\bea
\frac{d}{dY} \alpha^2_{r-s} &=&
- \frac{2g^2N_c}{(2\pi)^3} \int \frac{d^2z}{(r-z)^2}
  \left[\alpha_{r-s}^2 -\alpha_{z-s}^2  \right]
 +\frac{2g^2N_c}{(2\pi)^3} \int d^2z
  \frac{(s-z)\cdot(r-z)}{(s-z)^2 (r-z)^2}
  \left[\alpha_{z-z}^2 + \alpha_{r-s}^2 - 
        \alpha_{r-z}^2 - \alpha_{z-s}^2 \right]~. \label{eq:29}
\eea
To cast this into the familiar form for the BFKL equation we symmetrize the
kernel of the first term and rewrite that of the second term in a different
form:
\bea
\frac{2}{(r-z)^2} &=& \frac{1}{(r-z)^2} + \frac{1}{(s-z)^2} \\
2\frac{(s-z)\cdot(r-z)}{(s-z)^2 (r-z)^2} &=&
 \frac{1}{(r-z)^2} + \frac{1}{(s-z)^2} - \frac{(r-s)^2}{(s-z)^2 (r-z)^2}~.
\eea
Then, the part of~(\ref{eq:29}) involving $\alpha_{r-s}^2$ (the BFKL
virtual part) becomes
\be
-\frac{g^2N_c}{(2\pi)^3} \int d^2z
\frac{(r-s)^2}{(s-z)^2 (r-z)^2} \alpha_{r-s}^2
=
-\frac{\bar\alpha}{2\pi} \int d^2z
\frac{(r-s)^2}{(s-z)^2 (r-z)^2} \alpha_{r-s}^2~,
\ee
where $\bar\alpha\equiv\alpha_sN_c/\pi$.

The other terms from~(\ref{eq:29}) where $\alpha^2$ depends on the integration
variable $z$ (the BFKL real emissions) turn into
\bea
& & \frac{g^2N_c}{(2\pi)^3} \int d^2z \left\{
  \frac{\alpha_{z-s}^2}{(r-z)^2} + \frac{\alpha_{z-r}^2}{(s-z)^2}
 +\left[
    \frac{1}{(r-z)^2} + \frac{1}{(s-z)^2} - \frac{(r-s)^2}{(s-z)^2 (r-z)^2}
  \right]
  \left[\alpha_{z-z}^2 - \alpha_{r-z}^2 - \alpha_{s-z}^2 \right] \right\} \\
&=&
  \frac{\bar\alpha}{2\pi} \int d^2z \left\{
  -\frac{\alpha_{r-z}^2}{(r-z)^2} - \frac{\alpha_{s-z}^2}{(s-z)^2}
  +\frac{\alpha_0^2}{(r-z)^2} + \frac{\alpha_0^2}{(s-z)^2}
  +\frac{(r-s)^2}{(s-z)^2 (r-z)^2}
      \left(\alpha_{r-z}^2 + \alpha_{s-z}^2 - \alpha_0^2\right) \right\} \\
&=&
  \frac{\bar\alpha}{2\pi} \int d^2z \left\{
  -2\frac{\alpha_{z}^2 - \alpha_0^2}{z^2}
  +\frac{(r-s)^2}{(s-z)^2 (r-z)^2}
      \left(\alpha_{r-z}^2 + \alpha_{s-z}^2 - \alpha_0^2\right) \right\} ~.
\eea
%

\section{Four-point function}
The JIMWLK equation for the four-point function is
\bea
\frac{d}{dY} 
\langle \alpha_r^a \alpha_{\bar r}^b \alpha_s^c \alpha_{\bar s}^d \rangle &=&
  \frac{1}{2} \int d^2x d^2y \frac{\delta}{\delta\alpha_x^e}
     \eta^{ef}_{xy} \frac{\delta}{\delta\alpha_y^f} 
     \alpha_r^a \alpha_{\bar r}^b \alpha_s^c \alpha_{\bar s}^d \\
 &=& \frac{1}{2} \int d^2x d^2y \left[ \frac{\delta}{\delta\alpha_x^e}
     \eta^{ef}_{xy} \right] \frac{\delta}{\delta\alpha_y^f} 
     \alpha_r^a \alpha_{\bar r}^b \alpha_s^c \alpha_{\bar s}^d
  ~+~ \frac{1}{2} \int d^2x d^2y \, \eta^{ef}_{xy} 
     \frac{\delta^2}{\delta\alpha_x^e\, \delta\alpha_y^f} 
     \alpha_r^a \alpha_{\bar r}^b \alpha_s^c \alpha_{\bar s}^d ~, \label{eq:37}
\eea
We begin with the first term. The square bracket can be taken over from
eq.~(\ref{eq:5b}), and may also be written in the form
\bea
\frac{\delta}{\delta\alpha_x^e} \eta^{ef}_{xy} &=& 
- \frac{g}{\pi} f^{egf} \delta(x-y) \int \frac{d^2z}{(2\pi)^2}
  \frac{1}{(y-z)^2} \left[V^\dagger_z V_y \right]^{eg}~. 
\eea
We also need
\bea
\frac{\delta}{\delta\alpha_y^f} 
     \alpha_r^a \alpha_{\bar r}^b \alpha_s^c \alpha_{\bar s}^d &=&
\delta^{af}_{r-y} \alpha_{\bar r}^b \alpha_s^c \alpha_{\bar s}^d +
\delta^{bf}_{\bar{r}-y} \alpha_r^a \alpha_s^c \alpha_{\bar s}^d +
\delta^{cf}_{s-y} \alpha_r^a \alpha_{\bar r}^b \alpha_{\bar s}^d +
\delta^{df}_{\bar{s}-y} \alpha_r^a \alpha_{\bar r}^b \alpha_s^c ~.
\eea
The first term on the rhs of~(\ref{eq:37}) becomes
\bea
-\frac{g}{(2\pi)^3} \int d^2z & &
\left\{ 
  \frac{f^{ega}}{(r-z)^2} \left[V^\dagger_z V_r \right]^{eg}
  \alpha_{\bar r}^b \alpha_s^c \alpha_{\bar s}^d
+ \frac{f^{egb}}{(\bar{r}-z)^2} \left[V^\dagger_z V_{\bar r} \right]^{eg}
  \alpha_r^a \alpha_s^c \alpha_{\bar s}^d \right. \nonumber\\
& & \left.
+ \frac{f^{egc}}{(s-z)^2} \left[V^\dagger_z V_s \right]^{eg}
  \alpha_r^a \alpha_{\bar r}^b \alpha_{\bar s}^d
+ \frac{f^{egd}}{(\bar{s}-z)^2} \left[V^\dagger_z V_{\bar s} \right]^{eg}
  \alpha_r^a \alpha_{\bar r}^b \alpha_s^c
\right\} ~. \label{eq:40}
\eea
To linear order in the fields,
\be
f^{aeg} \left[V^\dagger_z V_r \right]^{eg} = 
ig f^{aeg} \left( \alpha_z^\kappa - \alpha_r^\kappa \right) 
           \left(t^\kappa\right)^{eg} =
-g f^{aeg} f^{\kappa eg} \left( \alpha_z^\kappa - \alpha_r^\kappa \right) =
-g N_c \left( \alpha_z^a - \alpha_r^a \right) ~,
\ee
and so eq.~(\ref{eq:40}) is equal to
\bea
& & \frac{g^2N_c}{(2\pi)^3} \int d^2z
\left\{ 
  \frac{1}{(r-z)^2} \left(\alpha_z^a - \alpha_r^a \right)
  \alpha_{\bar r}^b \alpha_s^c \alpha_{\bar s}^d
+ \frac{1}{(\bar{r}-z)^2} \alpha_r^a \left(\alpha_z^b - \alpha_{\bar r}^b \right)
  \alpha_s^c \alpha_{\bar s}^d \right. \nonumber\\
& & \hspace{2cm} \left.
+ \frac{1}{(s-z)^2} \alpha_r^a \alpha_{\bar r}^b 
  \left(\alpha_z^c - \alpha_s^c \right) \alpha_{\bar s}^d
+ \frac{1}{(\bar{s}-z)^2} \alpha_r^a \alpha_{\bar r}^b \alpha_s^c
  \left(\alpha_z^d - \alpha_{\bar s}^d \right)
\right\} \\
&=& 
-4 \frac{g^2N_c}{(2\pi)^3} \int \frac{d^2z}{z^2} \,
   \alpha_r^a \alpha_{\bar r}^b \alpha_s^c \alpha_{\bar s}^d \nonumber\\
& & + \frac{g^2N_c}{(2\pi)^3} \int d^2z
\left\{
  \frac{\alpha_z^a \alpha_{\bar r}^b \alpha_s^c \alpha_{\bar s}^d}{(r-z)^2} +
  \frac{\alpha_r^a \alpha_z^b \alpha_s^c \alpha_{\bar s}^d}{(\bar{r}-z)^2} +
  \frac{\alpha_r^a \alpha_{\bar r}^b \alpha_z^c \alpha_{\bar s}^d}{(s-z)^2} +
  \frac{\alpha_r^a \alpha_{\bar r}^b \alpha_s^c \alpha_z^d}{(\bar{s}-z)^2}
\right\} \label{eq:43}
\eea
This is the final result for the first term from~(\ref{eq:37}). In the
next section~\ref{sec:Gauss} we will simplify this further by taking the
expectation value with a Gaussian weight.

We now turn to the second term from eq.~(\ref{eq:37}). With $\eta^{ab}_{xy} 
= \eta^{ba}_{yx}$ we find
\be \label{eq:44}
\frac{1}{2} \int d^2x d^2y \, \eta^{ef}_{xy} 
     \frac{\delta^2}{\delta\alpha_x^e\, \delta\alpha_y^f} 
     \alpha_r^a \alpha_{\bar r}^b \alpha_s^c \alpha_{\bar s}^d =
 \eta^{ab}_{r\bar r} \alpha_s^c \alpha_{\bar s}^d
+\eta^{ac}_{rs} \alpha_{\bar r}^b \alpha_{\bar s}^d
+\eta^{ad}_{r\bar s} \alpha_{\bar r}^b \alpha_s^c
+\eta^{bc}_{\bar r s} \alpha_r^a \alpha_{\bar s}^d
+\eta^{bd}_{\bar r \bar s} \alpha_r^a \alpha_s^c
+\eta^{cd}_{s\bar s} \alpha_r^a \alpha_{\bar r}^b ~.
\ee
The expansion of $\eta^{ab}_{xy}$ in powers of the field starts
at second order. From~(\ref{eq:9}),
\bea
\eta^{ab}_{xy} &=& 
  \frac{g^2}{\pi} \int \frac{d^2z}{(2\pi)^2}
    \frac{(x-z)\cdot(y-z)}{(x-z)^2 (y-z)^2} 
    \left[\alpha_x\alpha_y - \alpha_x\alpha_z - \alpha_z\alpha_y +
          \alpha_z^2 + \alpha_z^2 \right]^{ab} \\
&=&
  \frac{g^2}{\pi} f^{e\kappa a}f^{f\kappa b} \int \frac{d^2z}{(2\pi)^2}
    \frac{(x-z)\cdot(y-z)}{(x-z)^2 (y-z)^2} 
    \left[\alpha_x^e\alpha_y^f - \alpha_x^e\alpha_z^f - \alpha_z^e\alpha_y^f +
          \alpha_z^e\alpha_z^f + \alpha_z^e\alpha_z^f \right] ~.
\label{eq:46}
\eea
We note that the two terms $\alpha_z^2$ exhibit different ordering in
rapidity as they arise from the expansion of $V_z$ and $V^\dagger_z$
to order $g^2$, respectively. For what follows, we combine them into a single
$\sim \alpha_z^2$ term which shows up no matter how rapidities are ordered.
Using~(\ref{eq:46}) in~(\ref{eq:44}) we obtain
\bea
\frac{g^2}{\pi} \int \frac{d^2z}{(2\pi)^2} & & \left\{
 f^{e\kappa a}f^{f\kappa b} \frac{(r-z)\cdot(\bar r-z)}{(r-z)^2 (\bar r-z)^2} 
 \left[\alpha_r^e\alpha_{\bar r}^f - \alpha_r^e\alpha_z^f - 
       \alpha_z^e\alpha_{\bar r}^f + \alpha_z^e\alpha_z^f \right]
       \alpha_s^c \alpha_{\bar s}^d \right. \nonumber\\
&+&
 f^{e\kappa a}f^{f\kappa c} \frac{(r-z)\cdot(s-z)}{(r-z)^2 (s-z)^2} 
 \left[\alpha_r^e\alpha_s^f - \alpha_r^e\alpha_z^f - 
       \alpha_z^e\alpha_s^f + \alpha_z^e\alpha_z^f \right]
       \alpha_{\bar r}^b \alpha_{\bar s}^d \nonumber\\
&+&
 f^{e\kappa a}f^{f\kappa d} \frac{(r-z)\cdot(\bar s-z)}{(r-z)^2 (\bar s-z)^2} 
 \left[\alpha_r^e\alpha_{\bar s}^f - \alpha_r^e\alpha_z^f - 
       \alpha_z^e\alpha_{\bar s}^f + \alpha_z^e\alpha_z^f \right]
       \alpha_{\bar r}^b \alpha_s^c \nonumber\\
&+&
 f^{e\kappa b}f^{f\kappa c} \frac{(\bar r-z)\cdot(s-z)}{(\bar r-z)^2 (s-z)^2} 
 \left[\alpha_{\bar r}^e\alpha_s^f - \alpha_{\bar r}^e\alpha_z^f - 
       \alpha_z^e\alpha_s^f + \alpha_z^e\alpha_z^f \right]
       \alpha_r^a \alpha_{\bar s}^d \nonumber\\
&+&
 f^{e\kappa b}f^{f\kappa d} \frac{(\bar r-z)\cdot(\bar s-z)}{(\bar r-z)^2 (\bar s-z)^2} 
 \left[\alpha_{\bar r}^e\alpha_{\bar s}^f - \alpha_{\bar r}^e\alpha_z^f - 
       \alpha_z^e\alpha_{\bar s}^f + \alpha_z^e\alpha_z^f \right]
       \alpha_r^a \alpha_s^c \nonumber\\
&+& \left.
 f^{e\kappa c}f^{f\kappa d} \frac{(s-z)\cdot(\bar s-z)}{(s-z)^2 (\bar s-z)^2} 
 \left[\alpha_s^e\alpha_{\bar s}^f - \alpha_s^e\alpha_z^f - 
       \alpha_z^e\alpha_{\bar s}^f + \alpha_z^e\alpha_z^f \right]
       \alpha_r^a \alpha_{\bar r}^b \right\}~. \label{eq:47a}
\eea
This is the final result for the second term from~(\ref{eq:37}). In
the next section we compute its expectation value for a Gaussian
weight.

\subsection{Gaussian approximation} \label{sec:Gauss}

To simplify the expressions further, we assume that the expectation
value of the four-point function on the rhs of the evolution equation
is taken with a Gaussian weight so that it is given by a sum over all
possible Wick contractions:
\bea
\langle \alpha_r^a \alpha_{\bar r}^b \alpha_s^c \alpha_{\bar s}^d \rangle
&=&
\langle \alpha_r^a \alpha_s^c \rangle \, 
   \langle \alpha_{\bar r}^b \alpha_{\bar s}^d \rangle +
\langle \alpha_r^a \alpha_{\bar r}^b \rangle \, 
   \langle \alpha_s^c \alpha_{\bar s}^d \rangle +
\langle \alpha_r^a \alpha_{\bar s}^d \rangle \, 
   \langle \alpha_{\bar r}^b \alpha_s^c \rangle \label{eq:A44}\\
&=&
\delta^{ac} \delta^{bd} \alpha^2_{r-s} \alpha^2_{\bar r-\bar s} +
\delta^{ab} \delta^{cd} \alpha^2_{r-\bar r} \alpha^2_{s-\bar s} +
\delta^{ad} \delta^{bc} \alpha^2_{r-\bar s} \alpha^2_{\bar r-s} ~,\label{eq:45}
\eea
where we used~(\ref{eq:28}). The above factorization into two-point
functions reproduces the evolution of the four-point function at
leading order in $N_c$. The lhs of~(\ref{eq:37}) then becomes
\bea
\frac{d}{dY} 
\langle \alpha_r^a \alpha_{\bar r}^b \alpha_s^c \alpha_{\bar s}^d \rangle &=&
\delta^{ac} \delta^{bd} \left(
    \alpha^2_{\bar r-\bar s} \frac{d}{dY} \alpha^2_{r-s} +
    \alpha^2_{r-s} \frac{d}{dY} \alpha^2_{\bar r-\bar s} \right) \label{eq:50}\\
& + &
\delta^{ab} \delta^{cd} \left(
    \alpha^2_{s-\bar s} \frac{d}{dY} \alpha^2_{r-\bar r} +
    \alpha^2_{r-\bar r} \frac{d}{dY} \alpha^2_{s-\bar s} \right) \label{eq:51}\\
& + &
\delta^{ad} \delta^{bc} \left(
    \alpha^2_{\bar r-s} \frac{d}{dY} \alpha^2_{r-\bar s} +
    \alpha^2_{r-\bar s} \frac{d}{dY} \alpha^2_{\bar r-s} \right) ~.
\label{eq:52}
\eea
We continue to analyze only one color channel corresponding to the first term
$\sim \delta^{ac} \delta^{bd}$ (the others are similar). The two-point
functions satisfy the BFKL equation and therefore
\bea
\alpha^2_{\bar r-\bar s} \frac{d}{dY} \alpha^2_{r-s} &=&
 -\frac{g^2 N_c}{(2\pi)^3} \int d^2z \left\{
    \frac{(r-s)^2}{(r-z)^2 (s-z)^2}
    \left[\alpha^2_{r-s} - \alpha^2_{r-z} - \alpha^2_{s-z} + {\alpha_0^2} 
\right]  + 2\frac{\alpha_{z}^2 - \alpha_0^2}{z^2} \right\}  \alpha^2_{\bar r-\bar s} \\
\alpha^2_{r-s} \frac{d}{dY} \alpha^2_{\bar r-\bar s} &=&
 -\frac{g^2 N_c}{(2\pi)^3} \int d^2z \left\{
    \frac{(\bar r-\bar s)^2}{(\bar r-z)^2 (\bar s-z)^2}
    \left[\alpha^2_{\bar r-\bar s} - \alpha^2_{\bar r-z} - \alpha^2_{\bar s-z} + {\alpha_0^2} 
\right] + 2\frac{\alpha_{z}^2 - \alpha_0^2}{z^2} \right\}  \alpha^2_{r-s} ~.
\eea
Thus, (\ref{eq:50}) turns into
\bea
\frac{d}{dY} \alpha^2_{r-s} \alpha^2_{\bar r-\bar s} &=& 
 -\frac{g^2 N_c}{(2\pi)^3} \int d^2z \left\{
    \left[\frac{(r-s)^2}{(r-z)^2 (s-z)^2} + 
          \frac{(\bar r-\bar s)^2}{(\bar r-z)^2 (\bar s-z)^2}
    \right]
    \alpha^2_{r-s} \alpha^2_{\bar r-\bar s} \right. \nonumber \\
& & +
  \left[ 
  2\frac{\alpha_{z}^2 - \alpha_0^2}{z^2}
    -\frac{(r-s)^2}{(r-z)^2 (s-z)^2}
    \left( \alpha^2_{r-z} + \alpha^2_{s-z} - \alpha_0^2 \right) 
  \right] \alpha^2_{\bar r-\bar s}
  \nonumber\\
& & \left. +
  \left[
2\frac{\alpha_{z}^2 - \alpha_0^2}{z^2}
    -\frac{(\bar r-\bar s)^2}{(\bar r-z)^2 (\bar s-z)^2}
    \left( \alpha^2_{\bar r-z} + \alpha^2_{\bar s-z} - \alpha_0^2 \right) 
  \right] \alpha^2_{r-s}
\right\} ~. \label{eq:55}
\eea
At leading order in $N_c$ the evolution of the four-point function
should be determined by the BFKL evolution of the two-point function,
as given in the previous equation [plus similar terms from
eqs.~(\ref{eq:51},\ref{eq:52})]. In what follows, we verify this
explicitly from the full JIMWLK evolution equation, eq.~(\ref{eq:43})
plus~(\ref{eq:47a}). However, we also derive the contributions which
are subleading in $N_c$.

\subsubsection{Virtual terms from JIMWLK}
Forming pairwise contractions on eq.~(\ref{eq:43}) gives
\bea
& &
\frac{g^2N_c}{(2\pi)^3} \int d^2z \biggl\{ \nonumber\\
& &
\delta^{ab} \delta^{cd} \left[
 -4 \frac{\alpha^2_{r-\bar r} \alpha^2_{s-\bar s}}{z^2} +
 \frac{\alpha_{z-\bar r}^2 \alpha_{s-\bar s}^2}{(r-z)^2} +
 \frac{\alpha_{r-z}^2 \alpha_{s-\bar s}^2 }{(\bar{r}-z)^2} +
 \frac{\alpha_{r-\bar r}^2 \alpha_{z-\bar s}^2}{(s-z)^2} +
 \frac{\alpha_{r-\bar r}^2 \alpha_{s-z}^2}{(\bar{s}-z)^2}
 \right] \nonumber\\
& &
\delta^{ac} \delta^{bd} \left[
 -4 \frac{\alpha^2_{r-s} \alpha^2_{\bar r-\bar s}}{z^2} +
 \frac{\alpha_{z-s}^2 \alpha_{\bar r-\bar s}^2}{(r-z)^2} +
 \frac{\alpha_{r-s}^2 \alpha_{z-\bar s}^2 }{(\bar{r}-z)^2} +
 \frac{\alpha_{r-z}^2 \alpha_{\bar r-\bar s}^2}{(s-z)^2} +
 \frac{\alpha_{r-s}^2 \alpha_{\bar r-z}^2}{(\bar{s}-z)^2}
 \right] \nonumber\\
& & \left.
\delta^{ad} \delta^{bc} \left[
 -4 \frac{\alpha^2_{r-\bar s} \alpha^2_{\bar r-s}}{z^2} +
 \frac{\alpha_{z-\bar s}^2 \alpha_{\bar r-s}^2}{(r-z)^2} +
 \frac{\alpha_{r-\bar s}^2 \alpha_{z-s}^2 }{(\bar{r}-z)^2} +
 \frac{\alpha_{r-\bar s}^2 \alpha_{\bar r-z}^2}{(s-z)^2} +
 \frac{\alpha_{r-z}^2 \alpha_{\bar r-s}^2}{(\bar{s}-z)^2}
 \right]
\right\}~.   \label{eq:53}
\eea
All of these terms will cancel against corresponding pieces
from eq.~(\ref{eq:47a}).

We organize the virtual contributions (where none of the fields depends
on the integration variable $z$) from eq.~(\ref{eq:47a}) according to the
various possible contractions. The term involving 
$\alpha^2_{r-s} \alpha^2_{\bar r-\bar s}$ is
\bea
& & \hspace{-2.2cm} \frac{g^2}{(2\pi)^3} \int d^2z \left\{
 f^{e\kappa a}f^{f\kappa b} \delta^{ec}\delta^{fd} 
 \left[\frac{1}{(r-z)^2}+\frac{1}{(\bar r-z)^2}-
   \frac{(r-\bar r)^2}{(r-z)^2 (\bar r-z)^2} \right]
  \right. \nonumber\\
&+&
 f^{e\kappa a}f^{f\kappa c} \delta^{ef}\delta^{bd} 
 \left[\frac{1}{(r-z)^2}+\frac{1}{(s-z)^2}-
   \frac{(r-s)^2}{(r-z)^2 (s-z)^2} \right] \nonumber\\
&+&
 f^{e\kappa a}f^{f\kappa d} \delta^{ec}\delta^{bf}
 \left[\frac{1}{(r-z)^2}+\frac{1}{(\bar s-z)^2}-
   \frac{(r-\bar s)^2}{(r-z)^2 (\bar s-z)^2} \right] \nonumber\\
&+&
 f^{e\kappa b}f^{f\kappa c} \delta^{af}\delta^{ed}
 \left[\frac{1}{(s-z)^2}+\frac{1}{(\bar r-z)^2}-
   \frac{(s-\bar r)^2}{(s-z)^2 (\bar r-z)^2} \right] \nonumber\\
&+&
 f^{e\kappa b}f^{f\kappa d} \delta^{ac}\delta^{ef}
 \left[\frac{1}{(\bar s-z)^2}+\frac{1}{(\bar r-z)^2}-
   \frac{(\bar s-\bar r)^2}{(\bar s-z)^2 (\bar r-z)^2} \right] \nonumber\\
&+& \left.
 f^{e\kappa c}f^{f\kappa d} \delta^{ae}\delta^{bf}
 \left[\frac{1}{(s-z)^2}+\frac{1}{(\bar s-z)^2}-
   \frac{(s-\bar s)^2}{(s-z)^2 (\bar s-z)^2} \right]
 \right\} \alpha^2_{r-s} \alpha^2_{\bar r-\bar s} \\
& & \hspace{-2.2cm} = \frac{g^2}{(2\pi)^3} \int d^2z \left\{
 f^{c\kappa a}f^{d\kappa b}
  \left[\frac{1}{(r-z)^2}+\frac{1}{(\bar r-z)^2}-
   \frac{(r-\bar r)^2}{(r-z)^2 (\bar r-z)^2} \right]
  \right. \nonumber\\
&+&
N_c \delta^{ac}\delta^{bd} 
 \left[\frac{1}{(r-z)^2}+\frac{1}{(s-z)^2}-
   \frac{(r-s)^2}{(r-z)^2 (s-z)^2} \right] \nonumber\\
&+&
 f^{c\kappa a}f^{b\kappa d}
 \left[\frac{1}{(r-z)^2}+\frac{1}{(\bar s-z)^2}-
   \frac{(r-\bar s)^2}{(r-z)^2 (\bar s-z)^2} \right] \nonumber\\
&+&
 f^{d\kappa b}f^{a\kappa c}
 \left[\frac{1}{(s-z)^2}+\frac{1}{(\bar r-z)^2}-
   \frac{(s-\bar r)^2}{(s-z)^2 (\bar r-z)^2} \right] \nonumber\\
&+&
 N_c \delta^{ac}\delta^{bd}
 \left[\frac{1}{(\bar s-z)^2}+\frac{1}{(\bar r-z)^2}-
   \frac{(\bar s-\bar r)^2}{(\bar s-z)^2 (\bar r-z)^2} \right] \nonumber\\
&+& \left.
 f^{a\kappa c}f^{b\kappa d}
 \left[\frac{1}{(s-z)^2}+\frac{1}{(\bar s-z)^2}-
   \frac{(s-\bar s)^2}{(s-z)^2 (\bar s-z)^2} \right]
 \right\} \alpha^2_{r-s} \alpha^2_{\bar r-\bar s} \\
& & \hspace{-2.2cm} = \frac{g^2}{(2\pi)^3} \int d^2z \left\{
 N_c \delta^{ac}\delta^{bd} 
 \left[\frac{4}{z^2}-
   \frac{(r-s)^2}{(r-z)^2 (s-z)^2} -
   \frac{(\bar r-\bar s)^2}{(\bar r-z)^2 (\bar s-z)^2}
 \right] 
\right. \nonumber\\
&-& \left.
 f^{a\kappa c}f^{b\kappa d}
 \left[
   \frac{(r-\bar r)^2}{(r-z)^2 (\bar r-z)^2} - 
   \frac{(r-\bar s)^2}{(r-z)^2 (\bar s-z)^2} -
   \frac{(s-\bar r)^2}{(s-z)^2 (\bar r-z)^2} + 
   \frac{(\bar s-s)^2}{(\bar s-z)^2 (s-z)^2}
 \right]
\right\} \alpha^2_{r-s} \alpha^2_{\bar r-\bar s} ~.  \label{eq:56}
\eea
The first term on the first line cancels the corresponding term involving
$\alpha^2_{r-s} \alpha^2_{\bar r-\bar s}$ from eq.~(\ref{eq:53}). The rest
of that line contributes to the rapidity evolution of the four-point
function at order ${\cal O}(N_c)$ and is seen to match the first line
of eq.~(\ref{eq:55}), i.e.\ the ``virtual correction'' part of the
leading-$N_c$ BFKL equation for the four-point function.
The second line in~(\ref{eq:56}) is a correction
which is suppressed by a relative factor of $1/N_c$.

Along the same lines one finds that the contribution from
eq.~(\ref{eq:47a}) involving $\alpha^2_{r-\bar r} \alpha^2_{s-\bar s}$
is
\bea
& & \hspace{-2.2cm} = \frac{g^2}{(2\pi)^3} \int d^2z \left\{
 N_c \delta^{ab}\delta^{cd} 
 \left[\frac{4}{z^2}-
   \frac{(r-\bar r)^2}{(r-z)^2 (\bar r-z)^2} -
   \frac{(s-\bar s)^2}{(s-z)^2 (\bar s-z)^2}
 \right] 
\right. \nonumber\\
&-& \left.
 f^{a\kappa b}f^{c\kappa d}
 \left[
   \frac{(r-s)^2}{(r-z)^2 (s-z)^2} - 
   \frac{(r-\bar s)^2}{(r-z)^2 (\bar s-z)^2} -
   \frac{(s-\bar r)^2}{(s-z)^2 (\bar r-z)^2} + 
   \frac{(\bar s-\bar r)^2}{(\bar s-z)^2 (\bar r-z)^2}
 \right]
\right\} \alpha^2_{r-\bar r} \alpha^2_{s-\bar s} ~.  \label{eq:57}
\eea
The first term on the first line will again cancel with a similar term 
$\sim -N_c\delta^{ab}\delta^{cd} \alpha^2_{r-\bar r} \alpha^2_{s-\bar s}$ 
from the first line of eq.~(\ref{eq:43}).

Finally, we also list the contribution from
eq.~(\ref{eq:47a}) involving $\alpha^2_{r-\bar s} \alpha^2_{\bar r - s}$:
\bea
& & \hspace{-2.2cm} = \frac{g^2}{(2\pi)^3} \int d^2z \left\{
 N_c \delta^{ad}\delta^{bc} 
 \left[\frac{4}{z^2}-
   \frac{(r-\bar s)^2}{(r-z)^2 (\bar s-z)^2} -
   \frac{(s-\bar r)^2}{(s-z)^2 (\bar r-z)^2}
 \right] 
\right. \nonumber\\
&-& \left.
 f^{a\kappa d}f^{b\kappa c}
 \left[
   \frac{(r-\bar r)^2}{(r-z)^2 (\bar r-z)^2} - 
   \frac{(r-s)^2}{(r-z)^2 (s-z)^2} -
   \frac{(\bar s-\bar r)^2}{(\bar s-z)^2 (\bar r-z)^2} + 
   \frac{(\bar s-s)^2}{(\bar s-z)^2 (s-z)^2}
 \right]
\right\} \alpha^2_{r-\bar s} \alpha^2_{\bar r - s} ~.\label{eq:58}
\eea
As before, the very first term in this expression will cancel. To
summarize the results obtained so far: the evolution equation for the
four-point function in mean-field approximation is given by
\bea
\frac{d}{dY} 
\langle \alpha_r^a \alpha_{\bar r}^b \alpha_s^c \alpha_{\bar s}^d \rangle &=&
\frac{d}{dY} \left(
   \delta^{ac} \delta^{bd} \alpha^2_{\bar r-\bar s} \alpha^2_{r-s} +
   \delta^{ab} \delta^{cd} \alpha^2_{s-\bar s} \alpha^2_{r-\bar r} +
   \delta^{ad} \delta^{bc} \alpha^2_{\bar r-s} \alpha^2_{r-\bar s}
   \right) \label{eq:62}\\
&-&  \frac{g^2}{(2\pi)^3} \int d^2z \Biggl\{ \nonumber\\
& &  f^{a\kappa c}f^{b\kappa d}
 \left[
   \frac{(r-\bar r)^2}{(r-z)^2 (\bar r-z)^2} - 
   \frac{(r-\bar s)^2}{(r-z)^2 (\bar s-z)^2} -
   \frac{(s-\bar r)^2}{(s-z)^2 (\bar r-z)^2} + 
   \frac{(\bar s-s)^2}{(\bar s-z)^2 (s-z)^2}
 \right] \alpha^2_{r-s} \alpha^2_{\bar r-\bar s} \nonumber\\
&+& f^{a\kappa b}f^{c\kappa d}
   \left[
    \frac{(r-s)^2}{(r-z)^2 (s-z)^2} - 
    \frac{(r-\bar s)^2}{(r-z)^2 (\bar s-z)^2} -
    \frac{(s-\bar r)^2}{(s-z)^2 (\bar r-z)^2} + 
    \frac{(\bar s-\bar r)^2}{(\bar s-z)^2 (\bar r-z)^2}
   \right] \alpha^2_{r-\bar r} \alpha^2_{s-\bar s} \nonumber\\
&+ & \left.
 f^{a\kappa d}f^{b\kappa c}
 \left[
   \frac{(r-\bar r)^2}{(r-z)^2 (\bar r-z)^2} - 
   \frac{(r-s)^2}{(r-z)^2 (s-z)^2} -
   \frac{(\bar s-\bar r)^2}{(\bar s-z)^2 (\bar r-z)^2} + 
   \frac{(\bar s-s)^2}{(\bar s-z)^2 (s-z)^2}
 \right] \alpha^2_{r-\bar s} \alpha^2_{\bar r - s}
\right\}~. \nonumber
\eea
The first line, eq.~(\ref{eq:62}), involves only standard BFKL
evolution of the two-point functions and provides the leading-$N_c$
contributions to both real emissions and virtual corrections. These
terms can be written in explicit form by using the BFKL equation for
the rapidity evolution for the two-point functions; for example, the
first term from that line is given in~(\ref{eq:55}). The remaining
lines correspond to the subleading (in $N_c$) contributions to the
virtual terms for JIMWLK evolution. In the next section, we derive the
corresponding real emission terms.

\subsubsection{Real terms from JIMWLK}

We begin with terms which contain only one $z$-dependent field,
as given by the second and third columns in~(\ref{eq:47a}). The first
line is given by
\bea
& & f^{c\kappa a}f^{f\kappa b} 
\left< \alpha_r^e \alpha_z^f \alpha_s^c \alpha_{\bar s}^d +
       \alpha_z^e \alpha_{\bar r}^f \alpha_s^c \alpha_{\bar s}^d
\right> \\
&=& f^{c\kappa a}f^{f\kappa b} \nonumber\\
& & \times
\left[ \delta^{ef} \delta^{cd} \left( \alpha^2_{r-z} \alpha^2_{s-\bar s} +
                                    \alpha^2_{z-\bar r} \alpha^2_{s-\bar s} 
                             \right) +
       \delta^{ec} \delta^{fd} \left( \alpha^2_{r-s} \alpha^2_{z-\bar s} +
                                    \alpha^2_{z-s} \alpha^2_{\bar r-\bar s} 
                             \right) +
       \delta^{ed} \delta^{cf} \left( \alpha^2_{r-\bar s} \alpha^2_{z-s} +
                                    \alpha^2_{z-\bar s} \alpha^2_{\bar r-s}
                             \right)
\right] \\
&=& N_c \delta^{ab} \delta^{cd} \left( \alpha^2_{r-z} \alpha^2_{s-\bar s} +
                                    \alpha^2_{z-\bar r} \alpha^2_{s-\bar s} 
                             \right)
    + f^{c\kappa a}f^{d\kappa b} \left( \alpha^2_{r-s} \alpha^2_{z-\bar s} +
                                    \alpha^2_{z-s} \alpha^2_{\bar r-\bar s} 
                             \right) 
    + f^{d\kappa a}f^{c\kappa b} \left( \alpha^2_{r-\bar s} \alpha^2_{z-s} +
                                    \alpha^2_{z-\bar s} \alpha^2_{\bar r-s}
                             \right) ~.
\eea
The second plus third terms from the first line of~(\ref{eq:47a}) thus becomes
\bea
&-& \frac{g^2}{(2\pi)^3} \int d^2z 
\left( \frac{1}{(r-z)^2} + \frac{1}{(\bar r-z)^2} - 
       \frac{(\bar r-r)^2}{(r-z)^2(\bar r-z)^2} \right)
 \nonumber\\
& & \times \left[
      N_c \delta^{ab} \delta^{cd} 
        (\alpha^2_{r-z} \alpha^2_{s-\bar s} + \alpha^2_{\bar r-z} \alpha^2_{s-\bar s} )
    + f^{c\kappa a}f^{d\kappa b}
        (\alpha^2_{r-s} \alpha^2_{z-\bar s} + \alpha^2_{\bar r-\bar s} \alpha^2_{z-s} )
    + f^{d\kappa a}f^{c\kappa b}
        (\alpha^2_{r-\bar s} \alpha^2_{z-s} + \alpha^2_{\bar r-s} \alpha^2_{z-\bar s} )
\right]~.
\eea
The remaining terms from~(\ref{eq:47a}) with one $z$-dependent field are 
easily obtained via permutations of color indices and coordinates. The full
result is
\bea
&-& \frac{g^2}{(2\pi)^3} \int d^2z  \nonumber\\
& & \left\{ \left( \frac{1}{(r-z)^2} + \frac{1}{(\bar r-z)^2} - 
       \frac{(\bar r-r)^2}{(r-z)^2(\bar r-z)^2} \right)
 \right. \nonumber\\
& & \times \left[
      N_c \delta^{ab} \delta^{cd} 
        (\alpha^2_{r-z} \alpha^2_{s-\bar s} + \alpha^2_{\bar r-z} \alpha^2_{s-\bar s} )
    + f^{c\kappa a}f^{d\kappa b}
        (\alpha^2_{r-s} \alpha^2_{z-\bar s} + \alpha^2_{\bar r-\bar s} \alpha^2_{z-s} )
    + f^{d\kappa a}f^{c\kappa b}
        (\alpha^2_{r-\bar s} \alpha^2_{z-s} + \alpha^2_{\bar r-s} \alpha^2_{z-\bar s} )
    \right]
 \nonumber\\
& & + 
   \left( \frac{1}{(r-z)^2} + \frac{1}{(s-z)^2} - 
       \frac{(s-r)^2}{(r-z)^2(s-z)^2} \right)
 \nonumber\\
& & \times \left[
      N_c \delta^{ac} \delta^{bd} 
        (\alpha^2_{r-z} \alpha^2_{\bar r-\bar s} + \alpha^2_{s-z} \alpha^2_{\bar r-\bar s} )
    + f^{b\kappa a}f^{d\kappa c}
        (\alpha^2_{r-\bar r} \alpha^2_{z-\bar s} + \alpha^2_{s-\bar s} \alpha^2_{z-\bar r} )
    + f^{d\kappa a}f^{b\kappa c}
        (\alpha^2_{r-\bar s} \alpha^2_{z-\bar r} + \alpha^2_{\bar r-s} \alpha^2_{z-\bar s} )
\right]
 \nonumber\\
& & + 
   \left( \frac{1}{(r-z)^2} + \frac{1}{(\bar s-z)^2} - 
       \frac{(\bar s-r)^2}{(r-z)^2(\bar s-z)^2} \right)
 \nonumber\\
& & \times \left[
      N_c \delta^{ad} \delta^{bc} 
        (\alpha^2_{r-z} \alpha^2_{\bar r-s} + \alpha^2_{\bar s-z} \alpha^2_{\bar r-s} )
    + f^{b\kappa a}f^{c\kappa d}
        (\alpha^2_{r-\bar r} \alpha^2_{z-s} + \alpha^2_{s-\bar s} \alpha^2_{z-\bar r} )
    + f^{c\kappa a}f^{b\kappa d}
        (\alpha^2_{r-s} \alpha^2_{z-\bar r} + \alpha^2_{\bar r-\bar s} \alpha^2_{z-s} )
\right]
 \nonumber\\
& & + 
   \left( \frac{1}{(\bar r-z)^2} + \frac{1}{(s-z)^2} - 
       \frac{(s-\bar r)^2}{(\bar r-z)^2(s-z)^2} \right)
 \nonumber\\
& & \times \left[
      N_c \delta^{ad} \delta^{bc}
        (\alpha^2_{\bar r-z} \alpha^2_{r-\bar s} + \alpha^2_{s-z} \alpha^2_{r-\bar s} )
    + f^{a\kappa b}f^{d\kappa c}
        (\alpha^2_{r-\bar r} \alpha^2_{z-\bar s} + \alpha^2_{s-\bar s} \alpha^2_{z-r} )
    + f^{d\kappa b}f^{a\kappa c}
        (\alpha^2_{\bar r-\bar s} \alpha^2_{z-r} + \alpha^2_{r-s} \alpha^2_{z-\bar s} )
\right]
 \nonumber\\
& & + 
   \left( \frac{1}{(\bar r-z)^2} + \frac{1}{(\bar s-z)^2} - 
       \frac{(\bar s-\bar r)^2}{(\bar r-z)^2(\bar s-z)^2} \right)
 \nonumber\\
& & \times \left[
      N_c \delta^{ac} \delta^{bd}
        (\alpha^2_{\bar r-z} \alpha^2_{r-s} + \alpha^2_{\bar s-z} \alpha^2_{r-s} )
    + f^{a\kappa b}f^{c\kappa d}
        (\alpha^2_{r-\bar r} \alpha^2_{z-s} + \alpha^2_{s-\bar s} \alpha^2_{z-r} )
    + f^{c\kappa b}f^{a\kappa d}
        (\alpha^2_{\bar r-s} \alpha^2_{z-r} + \alpha^2_{r-\bar s} \alpha^2_{z-s} )
\right]
 \nonumber\\
& & + 
   \left( \frac{1}{(s-z)^2} + \frac{1}{(\bar s-z)^2} - 
       \frac{(\bar s-s)^2}{(s-z)^2(\bar s-z)^2} \right)
 \nonumber\\
& & \left. \times \left[
      N_c \delta^{ab} \delta^{cd}
        (\alpha^2_{s-z} \alpha^2_{r-\bar r} + \alpha^2_{\bar s-z} \alpha^2_{r-\bar r} )
    + f^{a\kappa c}f^{b\kappa d}
        (\alpha^2_{r-s} \alpha^2_{z-\bar r} + \alpha^2_{\bar r-\bar s} \alpha^2_{z-r} )
    + f^{b\kappa c}f^{a\kappa d}
        (\alpha^2_{\bar r-s} \alpha^2_{z-r} + \alpha^2_{r-\bar s} \alpha^2_{z-\bar r} )
\right]
\right\}~.
\eea
Adding now the terms from eq.~(\ref{eq:53}) with one $z$-dependent field, and those
from eq.~(\ref{eq:47a}) with two $z$-dependent fields, leads
to the sum of the following two contributions; the terms proportional to $N_c$
are given by
\bea
&-& \frac{g^2N_c}{(2\pi)^3} \int d^2z  \biggl\{ \nonumber\\
& &   \delta^{ab} \delta^{cd}
      \left[ \left( 2\frac{\alpha^2_z-\alpha^2_0}{z^2} -
             \frac{(\bar r-r)^2}{(r-z)^2(\bar r-z)^2}
               \left(\alpha^2_{r-z}+\alpha^2_{\bar r-z} -\alpha^2_0 \right)
             \right) \alpha^2_{s-\bar s} \right.\nonumber\\
& & \qquad\quad \left. +   \left( 2\frac{\alpha^2_z-\alpha^2_0}{z^2} -
             \frac{(\bar s-s)^2}{(s-z)^2(\bar s-z)^2}
               \left(\alpha^2_{s-z}+\alpha^2_{\bar s-z} -\alpha^2_0 \right)
             \right)  \alpha^2_{r-\bar r} \right] \nonumber\\
& & + \delta^{ac} \delta^{bd}
      \left[ \left( 2\frac{\alpha^2_z-\alpha^2_0}{z^2} -
             \frac{(s-r)^2}{(r-z)^2(s-z)^2}
               \left(\alpha^2_{r-z}+\alpha^2_{s-z} -\alpha^2_0 \right)
             \right) \alpha^2_{\bar r-\bar s} \right.\nonumber\\
& & \qquad\quad \left. +   \left( 2\frac{\alpha^2_z-\alpha^2_0}{z^2} -
             \frac{(\bar s-\bar r)^2}{(\bar s-z)^2(\bar r-z)^2}
               \left(\alpha^2_{\bar r-z}+\alpha^2_{\bar s-z} -\alpha^2_0 \right)
             \right) \alpha^2_{r-s} \right] \nonumber\\
& & + \delta^{ad} \delta^{bc}
      \left[ \left( 2\frac{\alpha^2_z-\alpha^2_0}{z^2} -
             \frac{(\bar s-r)^2}{(r-z)^2(\bar s-z)^2}
               \left(\alpha^2_{r-z}+\alpha^2_{\bar s-z} -\alpha^2_0 \right)
             \right) \alpha^2_{\bar r-s} \right.\nonumber\\
& & \qquad\quad \left. \left. 
           +   \left( 2\frac{\alpha^2_z-\alpha^2_0}{z^2} -
             \frac{(s-\bar r)^2}{(s-z)^2(\bar r-z)^2}
               \left(\alpha^2_{\bar r-z}+\alpha^2_{s-z} -\alpha^2_0 \right)
             \right) \alpha^2_{r-\bar s} \right] \right\}
\eea
These provide the leading-$N_c$ contributions and correspond
to BFKL evolution of the product of two two-point functions; compare to
eq.~(\ref{eq:55}). The genuine JIMWLK contribution is
\bea
&-& \frac{g^2}{(2\pi)^3} \int d^2z  \biggl\{ \nonumber\\
& & f^{a\kappa b}f^{c\kappa d} \left[~
      \left( 
        \frac{1}{(r-z)^2} - \frac{(s-r)^2}{(r-z)^2(s-z)^2} 
        -\frac{1}{(\bar r-z)^2} + \frac{(s-\bar r)^2}{(\bar r-z)^2(s-z)^2} 
      \right)
       \alpha^2_{r-\bar r} \alpha^2_{z-\bar s} \right. \nonumber\\
& & \quad \quad \quad \quad +
      \left( 
        \frac{1}{(s-z)^2} - \frac{(s-r)^2}{(r-z)^2(s-z)^2} 
        - \frac{1}{(\bar s-z)^2} + \frac{(\bar s-r)^2}{(r-z)^2(\bar s-z)^2} 
      \right)
       \alpha^2_{s-\bar s} \alpha^2_{z-\bar r}    \nonumber\\
& & \quad \quad \quad \quad +
      \left( 
        \frac{1}{(\bar r-z)^2} - \frac{(\bar s-\bar r)^2}{(\bar r-z)^2(\bar s-z)^2} 
        -\frac{1}{(r-z)^2} + \frac{(\bar s-r)^2}{(r-z)^2(\bar s-z)^2} 
      \right)
        \alpha^2_{r-\bar r} \alpha^2_{z-s}   \nonumber\\
& & \quad \quad \quad \quad  +
      \left( 
        \frac{1}{(\bar s-z)^2} - \frac{(\bar s-\bar r)^2}{(\bar r-z)^2(\bar s-z)^2} 
        - \frac{1}{(s-z)^2} + \frac{(s-\bar r)^2}{(\bar r-z)^2(s-z)^2} 
      \right)
        \alpha^2_{s-\bar s} \alpha^2_{z-r}  \nonumber\\
& & \quad \quad \quad \quad  +
 \left(  \frac{(s-r)^2}{(r-z)^2(s-z)^2} -  \frac{1}{(r-z)^2} -  \frac{1}{(s-z)^2}  \right)
     \alpha^2_{z-\bar s} \alpha^2_{z-\bar r}   \nonumber\\
& & \quad \quad \quad \quad  + 
 \left(  - \frac{(\bar s-r)^2}{(r-z)^2(s-z)^2} +  \frac{1}{(r-z)^2} +  \frac{1}{(\bar s-z)^2}  \right)
     \alpha^2_{z- s} \alpha^2_{z-\bar r}   \nonumber\\
& & \quad \quad \quad \quad  + 
 \left(  - \frac{(\bar r-s)^2}{(r-z)^2(s-z)^2} +  \frac{1}{(\bar r-z)^2} +  \frac{1}{( s-z)^2}  \right)
     \alpha^2_{z- \bar s} \alpha^2_{z- r}  \nonumber\\
& & \quad \quad \quad \quad  +
 \left. \left(  \frac{(\bar s-\bar r)^2}{(\bar r-z)^2(\bar s-z)^2} - \frac{1}{(\bar r-z)^2} - \frac{1}{(\bar s-z)^2}  \right)
     \alpha^2_{z- s} \alpha^2_{z- r} \right]\nonumber\\
&+& f^{a\kappa c}f^{b\kappa d} \left[~
      \left( 
        \frac{1}{(r-z)^2} - \frac{(\bar r-r)^2}{(r-z)^2(\bar r-z)^2} 
        - \frac{1}{(s-z)^2} + \frac{(s-\bar r)^2}{(\bar r-z)^2(s-z)^2} 
      \right)
       \alpha^2_{r-s} \alpha^2_{z-\bar s}  \right. \nonumber\\
& & \quad \quad \quad \quad +
      \left( 
        \frac{1}{(\bar r-z)^2} - \frac{(\bar r-r)^2}{(r-z)^2(\bar r-z)^2} 
        - \frac{1}{(\bar s-z)^2} + \frac{(\bar s-r)^2}{(r-z)^2(\bar s-z)^2} 
      \right)
       \alpha^2_{\bar r-\bar s} \alpha^2_{z-s}  \nonumber\\
& & \quad \quad \quad \quad +
      \left( 
        \frac{1}{(s-z)^2} - \frac{(\bar s-s)^2}{(s-z)^2(\bar s-z)^2} 
        -\frac{1}{(r-z)^2} + \frac{(\bar s-r)^2}{(r-z)^2(\bar s-z)^2} 
      \right)
        \alpha^2_{r-s} \alpha^2_{z-\bar r}  \nonumber\\
& & \quad \quad \quad \quad  +
      \left( 
        \frac{1}{(\bar s-z)^2} - \frac{(\bar s-s)^2}{(s-z)^2(\bar s-z)^2} 
        -\frac{1}{(\bar r-z)^2} + \frac{(s-\bar r)^2}{(\bar r-z)^2(s-z)^2} 
      \right)
        \alpha^2_{\bar r-\bar s} \alpha^2_{z-r}  \nonumber\\
& & \quad \quad \quad \quad  +
 \left(  \frac{(\bar s- s)^2}{(s-z)^2(\bar s-z)^2} -  \frac{1}{(\bar s -z)^2} -  \frac{1}{(s-z)^2}  \right)
     \alpha^2_{z-\bar r} \alpha^2_{z- r}   \nonumber\\
& & \quad \quad \quad \quad  + 
 \left(  - \frac{(\bar r-s)^2}{(\bar r-z)^2(s-z)^2} +  \frac{1}{(\bar r-z)^2} +  \frac{1}{(s-z)^2}  \right)
     \alpha^2_{z- \bar s} \alpha^2_{z- r}   \nonumber\\
& & \quad \quad \quad \quad  + 
 \left(  - \frac{(\bar s-r)^2}{(r-z)^2(\bar s-z)^2} +  \frac{1}{(r-z)^2} +  \frac{1}{(\bar s-z)^2}  \right)
     \alpha^2_{z- s} \alpha^2_{z- \bar r}  \nonumber\\
& & \quad \quad \quad \quad  +
 \left. \left(  \frac{(r-\bar r)^2}{(r-z)^2(\bar r-z)^2} - \frac{1}{(r-z)^2} - \frac{1}{(\bar r-z)^2}  \right)
     \alpha^2_{z- s} \alpha^2_{z- \bar s} \right]\nonumber\\
%
%
& & f^{a\kappa d}f^{b\kappa c} \left[~
      \left( 
        \frac{1}{(r-z)^2} - \frac{(\bar r-r)^2}{(r-z)^2(\bar r-z)^2} 
        - \frac{1}{(\bar s-z)^2} + \frac{(\bar s-\bar r)^2}{(\bar r-z)^2(\bar s-z)^2} 
      \right)
        \alpha^2_{r-\bar s} \alpha^2_{z-s} \right. \nonumber\\
& & \quad \quad \quad \quad +
      \left( 
        \frac{1}{(\bar r-z)^2} - \frac{(\bar r-r)^2}{(r-z)^2(\bar r-z)^2} 
        - \frac{1}{(s-z)^2} + \frac{(s-r)^2}{(r-z)^2(s-z)^2} 
      \right)
        \alpha^2_{\bar r-s} \alpha^2_{z-\bar s} \nonumber\\
& & \quad \quad \quad \quad +
      \left( 
        \frac{1}{(\bar s-z)^2} - \frac{(\bar s-s)^2}{(s-z)^2(\bar s-z)^2} 
        -\frac{1}{(r-z)^2} + \frac{(s-r)^2}{(r-z)^2(s-z)^2} 
      \right)
        \alpha^2_{r-\bar s} \alpha^2_{z-\bar r} \nonumber\\
& & \quad \quad \quad \quad  +
      \left( 
        \frac{1}{(s-z)^2} - \frac{(\bar s-s)^2}{(s-z)^2(\bar s-z)^2} 
        -\frac{1}{(\bar r-z)^2} + \frac{(\bar s-\bar r)^2}{(\bar r-z)^2(\bar s-z)^2} 
      \right)
        \alpha^2_{\bar r-s} \alpha^2_{z-r}  \nonumber\\
& & \quad \quad \quad \quad  +
 \left(  \frac{(\bar r- r)^2}{(r-z)^2(\bar r-z)^2} -  \frac{1}{(\bar r -z)^2} -  \frac{1}{(r-z)^2}  \right)
     \alpha^2_{z-\bar s} \alpha^2_{z- s}   \nonumber\\
& & \quad \quad \quad \quad  + 
 \left(  - \frac{(r-s)^2}{(r-z)^2 (s-z)^2} +  \frac{1}{(r-z)^2} +  \frac{1}{(s-z)^2}  \right)
     \alpha^2_{z- \bar r} \alpha^2_{z- \bar s}   \nonumber\\
& & \quad \quad \quad \quad  + 
 \left(  - \frac{(\bar s-\bar r)^2}{(\bar r-z)^2(\bar s-z)^2} +  \frac{1}{(\bar r-z)^2} +  \frac{1}{(\bar s-z)^2}  \right)
     \alpha^2_{z- s} \alpha^2_{z- r}  \nonumber\\
& & \quad \quad \quad \quad  +
 \left .\left. \left(  \frac{(s-\bar s)^2}{(s-z)^2(\bar s-z)^2} - \frac{1}{(s-z)^2} - \frac{1}{(\bar s-z)^2}  \right)
     \alpha^2_{z- r} \alpha^2_{z- \bar r}
\right] \right\}~.  \label{eq:70}
\eea
This expression~(\ref{eq:70}), which is to be added to the rhs of~(\ref{eq:62}), is our
final result for the evolution equation of the JIMWLK four-point function for a
Gaussian weight. It can be rewritten in more compact form by exploiting some of the
symmetries of $\langle\alpha_r^a \alpha_{\bar r}^b \alpha_s^c \alpha_{\bar s}^d \rangle$, eqs.~(\ref{eq:fact_correction},\ref{eq:Fs}).

\section{Color factors} \label{sec:ColorFactorsApp}

In this appendix, we compute the color factors for the products of a
leading-$N_c$ term from the first line~(\ref{eq:4point_leading}) with
a subleading-$N_c$ term from the second
line~(\ref{eq:4point_subleading}); color indices from the target side
carry a prime and we need to also include the remaining structure
constants from eq.~(\ref{eq:Cpq_FourPoint}). Using the following
SU($N_c$) identities
\bea
f^{ab\kappa}f^{cb\kappa} &=& N_c \, \delta^{ac}~,\\
f^{ab\kappa}f^{cd\kappa} &=& \frac{2}{N_c}\left(\delta^{ac}\delta^{bd}
- \delta^{ad}\delta^{bc}\right) 
        + d^{ac\kappa}d^{bd\kappa} - d^{ad\kappa}d^{bc\kappa}~,\\
d^{ac\kappa}d^{acr} &=& \left(N_c-\frac{4}{N_c}\right)\delta^{r\kappa}~,\\
f^{g aa^\prime} f^{gcc^\prime} f^{ac\kappa} =
i^3 \, {\rm tr_{adj}}\, t^{a^\prime} t^{c^\prime} t^{\kappa}  &=&
\frac{N_c}{2} f^{a' c' \kappa}~, \\
f^a_{a' g} f^c_{gc'} f^d_{c'g'} f^b_{g'a'} = {\rm tr_{adj}}\, t^a t^c t^d t^b
 &=& \delta^{ac}\delta^{bd} + \delta^{ab}\delta^{cd} + \frac{N_c}{4}
      \left(d^{acr}d^{dbr} - d^{adr}d^{bcr} + d^{abr}d^{cdr}\right)~,   \\
d^{ad\kappa}d^{bc\kappa}d^{acr}d^{bdr} &=& {\cal O}(N_c^4)~,
\eea
one derives
\bea
\frac{1}{N_c} \delta^{a'b'} \delta^{c'd'} 
  \left(f_{g aa^\prime} f_{gcc^\prime} f^{ac\kappa}\right)
  \left(f_{g^\prime b b^\prime} f_{g^\prime dd^\prime} f^{bd\kappa}\right) &=&
  \frac{1}{4}N_c^2 (N_c^2-1)~,\\
\frac{1}{N_c} \delta^{a'b'} \delta^{c'd'}
  f_{g aa^\prime} f_{gcc^\prime} f_{g^\prime b b^\prime} f_{g^\prime dd^\prime} f^{ab\kappa}
  f^{cd\kappa} &=& 
  \frac{1}{N_c} f^{ab\kappa} f^{cd\kappa} \, {\rm tr_{adj}}\, t^a t^c t^d t^b
      \nonumber\\
  = \frac{1}{N_c} f^{ab\kappa} f^{cd\kappa} \left(
      \delta^{ac}\delta^{bd} + \delta^{ab}\delta^{cd} + \frac{N_c}{4}
      \left(d^{acr}d^{dbr} - d^{adr}d^{bcr} + d^{abr}d^{cdr}\right)\right)
   &=&  \delta^{bd} + \frac{1}{2} f^{ab\kappa} f^{cd\kappa}d^{acr}d^{bdr} \nonumber
      \\ 
   & & \hspace{-6cm}= \left[1 - \frac{1}{N_c} \left(N_c-\frac{4}{N_c}\right)
      +\frac{1}{2} \left(N_c-\frac{4}{N_c}\right)^2
    +{\cal O}(N_c^2) \right] (N_c^2-1) \\
\frac{1}{N_c} \delta^{a'b'} \delta^{c'd'}
  f_{g aa^\prime} f_{gcc^\prime} f_{g^\prime b b^\prime} f_{g^\prime dd^\prime} 
  f^{ad\kappa} f^{bc\kappa} &=&
  \frac{1}{N_c} f^{ad\kappa} f^{bc\kappa} \, {\rm tr_{adj}}\, t^a t^c t^d t^b
      \nonumber\\
  = \frac{1}{N_c} f^{ad\kappa} f^{bc\kappa} \left(
      \delta^{ac}\delta^{bd} + \delta^{ab}\delta^{cd} + \frac{N_c}{4}
      \left(d^{acr}d^{dbr} - d^{adr}d^{bcr} + d^{abr}d^{cdr}\right)\right)
   &=& 0 \label{eq:C9} \\
\frac{1}{N_c} \delta^{a'c'} \delta^{b'd'}
  f_{g aa^\prime} f_{g^\prime b b^\prime} f_{gcc^\prime} f_{g^\prime dd^\prime}\,
  f^{ac\kappa}f^{bd\kappa} = N_c \delta^{ac} \delta^{bd}
  f^{ac\kappa}f^{bd\kappa} &=& 0 \\
\frac{1}{N_c} \delta^{a'c'} \delta^{b'd'} 
  f_{g aa^\prime} f_{g^\prime b b^\prime} f_{gcc^\prime} f_{g^\prime dd^\prime}\,
  f^{ab\kappa}f^{cd\kappa} =
  N_c \delta^{ac} \delta^{bd}\, f^{ab\kappa}f^{cd\kappa} &=& N_c^2(N_c^2-1) \\
\frac{1}{N_c} \delta^{a'c'} \delta^{b'd'}
  f_{g aa^\prime} f_{g^\prime b b^\prime} f_{gcc^\prime} f_{g^\prime dd^\prime}\,
  f^{ad\kappa}f^{bc\kappa} = N_c \delta^{ac} \delta^{bd}
  f^{ad\kappa}f^{bc\kappa} &=& - N_c^2 (N_c^2-1) \\
\frac{1}{N_c} \delta^{a'd'} \delta^{b'c'} 
  \left(f_{g aa^\prime} f_{gcc^\prime} f^{ac\kappa} \right)
  \left(f_{g^\prime b b^\prime} f_{g^\prime dd^\prime} f^{bd\kappa} \right) &=&
  - \frac{1}{4}N_c^2 (N_c^2-1)~,\\
\frac{1}{N_c} \delta^{a'd'} \delta^{b'c'} 
  f_{g aa^\prime} f_{g^\prime b b^\prime} f_{gcc^\prime} f_{g^\prime dd^\prime}\,
  f^{ab\kappa}f^{cd\kappa} &=&   
  \frac{1}{N_c} f^{ab\kappa} f^{cd\kappa} \, {\rm tr_{adj}}\, t^a t^c t^b t^d
      \nonumber\\
&=& - \mathrm{(\ref{eq:C9})} = 0 \\
\frac{1}{N_c} \delta^{a'd'} \delta^{b'c'} 
  f_{g aa^\prime} f_{g^\prime b b^\prime} f_{gcc^\prime} f_{g^\prime dd^\prime}\,
  f^{ad\kappa}f^{bc\kappa} &=& - \frac{1}{4}N_c^2 (N_c^2-1)~.
\eea

\begin{acknowledgments}
  We have used the JaxoDraw program~\cite{Binosi:2008ig} to draw the
  diagrams shown in this paper. J.J-M. thanks A. Kovner for useful discussions. 
We gratefully acknowledge support by
  the DOE Office of Nuclear Physics through Grant
  No.\ DE-FG02-09ER41620 and from The City University of New York
  through the PSC-CUNY Research Award Program, grants 60060-3940 (A.D.) and
  62625-40 (J.J.M.).
\end{acknowledgments}


\end{document}